\documentclass[manuscript, nonacm, 10pt]{acmart}
\definecolor{paleyellow}{rgb}{1, 1, 0.85}
\definecolor{lower}{rgb}{0.88235,0.7451,0.41569}
\definecolor{lower}{rgb}{0.94, 0.85, 0.6}
\definecolor{higher}{rgb}{0.5, 0.8, 0.75}
\definecolor{rowColor}{rgb}{0.937, 0.937, 0.937}
\definecolor{mauve}{rgb}{0.25,0,0.52}

\newcommand{\subheading}[1]{\vspace{2pt}\noindent\textbf{#1}}

\newcommand{\qualquote}[1]{%
  \textit{\textcolor{mauve}{``#1''}}
}

\newcommand{\summarytcolorbox}[1]{
\begin{tcolorbox}[colback=paleyellow, colframe=black, colbacktitle=white, coltitle=black, boxsep=0pt, top=3pt, bottom=3pt, left=6pt, right=6pt, before skip=3pt]
    \noindent #1
\end{tcolorbox}
}

\makeatletter
\patchcmd{\@makecaption}
  {\scshape}
  {}
  {}
  {}
\long\def\@makecaption#1#2{%
  \vskip\abovecaptionskip
  \small % Set caption font size to small
  \sbox\@tempboxa{#1. #2}%
  \ifdim \wd\@tempboxa >\hsize
    #1. #2\par
  \else
    \global \@minipagefalse
    \hb@xt@\hsize{\hfil\box\@tempboxa\hfil}%
  \fi
  \vskip\belowcaptionskip}
\makeatother

\definecolor{jennacolor}{RGB}{200, 50, 110}

\newcommand{\genstrat}[2]{(\textbf{S{#1}}, ${#2}\%$)}
\newcommand{\aistrat}[2]{(\textbf{AIS{#1}}, ${#2}\%$)}
\newcommand{\attentionstrat}[2]{(\textbf{AS{#1}}, ${#2}\%$)}

\usepackage{xcolor}
\usepackage{xspace}
\usepackage{tcolorbox}
\usepackage{fontawesome}
\usepackage{pgfplots}
\usepackage{pgfplotstable}
\usepackage{multirow}
\usepackage{multicol}

\usepackage{xcolor}
\usepackage{array}
\usepackage{booktabs}
\usepackage{tikz}

\definecolor{KaiaFreqOne}{HTML}{F3E8FF}
\definecolor{KaiaFreqTwo}{HTML}{D8C2F0}
\definecolor{KaiaFreqThree}{HTML}{B78AE6}
\definecolor{KaiaFreqFour}{HTML}{8D5BC7}
\definecolor{KaiaFreqFive}{HTML}{4C2D73}
\definecolor{KaiaCompOne}{HTML}{5E3C99}
\definecolor{KaiaCompTwo}{HTML}{B2ABD2}
\definecolor{KaiaCompThree}{HTML}{E6E6E6}
\definecolor{KaiaCompFour}{HTML}{FEE08B}
\definecolor{KaiaCompFive}{HTML}{E6AB02}

\newlength{\KaiaBarWidth}
\newlength{\KaiaEndWidth}
\newsavebox{\KaiaTableBox}

\newlength{\KaiaBarRaise}
\newlength{\KaiaHeaderKern}
\newcommand{\KaiaDistributionBar}[7]{%
  \begingroup
  \setlength{\KaiaBarWidth}{0.62\linewidth}%
  \setlength{\KaiaEndWidth}{0.18\linewidth}%
  \makebox[\linewidth][l]{%
    \if\relax\detokenize{#6}\relax
    \else
    \makebox[\KaiaEndWidth][r]{{\fontsize{6.5}{7.5}\selectfont\itshape #6\%}}%
    \hspace{0.01\linewidth}%
    \fi
    \raisebox{\KaiaBarRaise}{%
    \begin{tikzpicture}[x=\KaiaBarWidth,y=6.5pt,baseline=(current bounding box.center)]
      \path[use as bounding box] (0,0) rectangle (1,1);
      \clip (0,0) rectangle (1,1);
      \pgfmathsetmacro{\KaiaA}{#1}
      \pgfmathsetmacro{\KaiaB}{#1+#2}
      \pgfmathsetmacro{\KaiaC}{#1+#2+#3}
      \pgfmathsetmacro{\KaiaD}{#1+#2+#3+#4}
      \pgfmathsetmacro{\KaiaE}{#1+#2+#3+#4+#5}
      \fill[\KaiaColorOne] (0,0) rectangle (\KaiaA,1);
      \fill[\KaiaColorTwo] (\KaiaA,0) rectangle (\KaiaB,1);
      \fill[\KaiaColorThree] (\KaiaB,0) rectangle (\KaiaC,1);
      \fill[\KaiaColorFour] (\KaiaC,0) rectangle (\KaiaD,1);
      \fill[\KaiaColorFive] (\KaiaD,0) rectangle (\KaiaE,1);
      \foreach \boundary in {\KaiaA,\KaiaB,\KaiaC,\KaiaD}
        \draw[white,line width=0.18pt] (\boundary,0) -- (\boundary,1);
      \draw[black!25,line width=0.18pt] (0,0) rectangle (1,1);
    \end{tikzpicture}%
    }%
    \if\relax\detokenize{#7}\relax
    \else
      \hspace{0.01\linewidth}%
      \makebox[\KaiaEndWidth][l]{{\fontsize{6.5}{7.5}\selectfont\itshape #7\%}}%
    \fi
  }%
  \endgroup
}

\newlength{\KaiaPctWidth}   % add near \newlength{\KaiaEndWidth} if not already there

\newcommand{\KaiaDistributionHeader}[1]{%
\begingroup
\setlength{\KaiaBarWidth}{0.62\linewidth}%
\setlength{\KaiaEndWidth}{0.18\linewidth}%
\settowidth{\KaiaPctWidth}{{\fontsize{6.5}{7.5}\selectfont\itshape 88\%}}%
\makebox[\linewidth][l]{%
\hspace*{\dimexpr\KaiaEndWidth-\KaiaPctWidth+\KaiaHeaderKern\relax}%
\makebox[\KaiaPctWidth][l]{\textbf{#1}}%
\hspace{0.01\linewidth}%
\makebox[\KaiaBarWidth][l]{}%
\hspace{0.01\linewidth}%
\makebox[\KaiaEndWidth][l]{}%
}%
\endgroup
}

\newcommand{\KaiaFrequencyBar}[7]{%
  \begingroup
  \def\KaiaColorOne{KaiaFreqOne}%
  \def\KaiaColorTwo{KaiaFreqTwo}%
  \def\KaiaColorThree{KaiaFreqThree}%
  \def\KaiaColorFour{KaiaFreqFour}%
  \def\KaiaColorFive{KaiaFreqFive}%
  \KaiaDistributionBar{#1}{#2}{#3}{#4}{#5}{#6}{#7}%
  \endgroup
}

\newcommand{\KaiaLegend}[2]{%
  \raisebox{0.3pt}{\color{#1}\rule{5.5pt}{5.5pt}}\hspace{2pt}#2%
}

\newcommand{\KaiaCheckTableWidth}{%
  \typeout{KAIA_TABLE_WIDTH=\the\wd\KaiaTableBox}%
  \typeout{KAIA_COLUMN_WIDTH=\the\columnwidth}%
  \ifdim\wd\KaiaTableBox>\columnwidth
    \PackageError{kaia-tables}{Table exceeds ACM column width}{Do not use table* or scale the complete table.}%
  \fi
}
\AtBeginDocument{%
  }

\acmISBN{978-1-4503-XXXX-X/2018/06}

\usepackage{booktabs}
\usepackage{tabularx}
\usepackage{utfsym}
\usepackage[table]{xcolor}
\begin{document}
%%
%% The "title" command has an optional parameter, % allowing the author to
%define a "short title" to be used in page headers.
\title{Level Up! AI-Driven Challenges and Adaptations to Code Review Practices}
% ``Vicious and Dangerous Loop''

%%
%% The "author" command and its associated commands are used to define % the
%authors and their affiliations. % Of note is the shared affiliation of the
%first two authors, and the % "authornote" and "authornotemark" commands % used
%to denote shared contribution to the research.

\author{Kaia Newman} 
\authornote{Work was completed during an internship at Microsoft.}
\affiliation{%
  \institution{Carnegie Mellon University}
  \city{Pittsburgh}
  \country{USA}}
\email{kaian@cmu.edu}

\author{Christian Bird}
\affiliation{%
  \institution{Microsoft}
  \city{Redmond}
  \country{USA}
}
\email{cabird@gmail.com}

\author{Jenna Butler}
\affiliation{%
  \institution{Microsoft}
  \city{Redmond}
  \country{USA}
}
\email{jennbu@microsoft.com}

\author{Emerson Murphy-Hill}
\affiliation{%
  \institution{Microsoft}
  \city{Redmond}
  \country{USA}
}
\email{captain.emerson@gmail.com}

%%
%% By default, the full list of authors will be used in the page % headers.
%Often, this list is too long, and will overlap % other information printed in
%the page headers. This command allows % the author to define a more concise
%list % of authors' names for this purpose.
\renewcommand{\shortauthors}{Newman et al.}

%%
%% The abstract is a short summary of the work to be presented in the % article.
\begin{abstract}
AI can write code faster than humans can review, causing a bottleneck and
necessitating changes to code review practices. This bottleneck is hotly
debated, giving rise to two extreme and orthogonal imagined futures: one where
humans continue to be firmly entrenched in the review process, and one where AI
agents merge pull requests entirely without human involvement. Yet, these
futures remain rooted in folk wisdom. Through a mixed-methods study of
interviews ($n = 24$) and a survey ($n = 2{,}590$)
at a large, multinational software company with widespread access to AI, we identify AI-driven code review
challenges and strategies reported as effective to meet them.
We found that AI enables, necessitates, and complicates a \textbf{shift in the human reviewer's role toward high-level concerns} (e.g., design and intent).
We found that AI-assisted authoring results in \textbf{increased burden} for reviewers due to challenges such as a lack of author stewardship and PR volume and size, potentially leading to downstream debt. Our survey showed AI review usage
was \textit{not} correlated with less difficulty keeping up with PR volume,
indicating \textbf{using AI for review alone is not sufficient} for solving AI-driven review challenges.
Developers found the most useful strategies to be organizations supporting and resourcing review, pushing review burden upstream, and prioritizing reviewer attention on the concerns that matter most.

\end{abstract}

% the below is a solution to the blurring of the roles Developers indicated that
% the role of the human reviewer is shifting toward high-level concerns such as
% intent and architecture, and the most useful adaptations involve pushing
% review burden upstream to enable this level of review. 

%%
%% The code below is generated by the tool at http://dl.acm.org/ccs.cfm. %
%Please copy and paste the code instead of the example below.
%%




%%
%% This command processes the author and affiliation and title % information and
%builds the first part of the formatted document.
\maketitle

% Other title ideas: Code Review Isn't Dead; Authoring Is
\section{Introduction}
% \fixme{not sure how to cause reader to probe weakly-held belief up here? want
% provocative question} \textit{In a world where code is built by AI and
% verified by human authors, what role, if any, does a reviewer have?}

Across organizations, AI-assisted authoring has spurred  increases in \textit{pull request} (PR) volume that outpace human reviewing capacity, necessitating changes to code review practices~\cite{faros2025aiproductivity, he2026ai, li2025rise, appajodu2026scaling, googlecloud_ai_writes_code}. 
For example, at one mid-sized company, 
% At one mid-sized,
% AI-forward company, 
a top-down mandate to double throughput with AI catalyzed a symmetric increase in review load, pushing AI review to largely replace human review~\cite{he2026ai}. 
As PR volume outpaces review capacity, review scrutiny may decline and latency may increase~\cite{yu2026habituation, adams2026automating, song2025impact}.
Stakeholders await the return on investment from AI adoption, and code review
appears to be a culprit:
one study 
% looking at multidimensional measures of
% developer productivity 
% (e.g., performance and efficiency and flow) 
found that 
AI
adoption caused a systematic redistribution of developer effort rather
than increasing productivity, offsetting throughput with review
burden~\cite{afroz2025fast}. 

% However, this research sometimes reports conflicting
% results on important quantities (e.g., merge rates:~\cite{popescu2026investigating}, cf.~\cite{pham2026code}) and can admit
% opposite conclusions based on equally defensible methodological choices without
% a deeper understanding of why phenomena occur~\cite{}. 

% code review could be preventing us from realizing ai-promised productivity

% unsustainable to review using standard practices, often using AI to supplement 
% but I also like hot debate. how to keep both?

% new para --> automated review is a natural consequence/solution: consequence in that AI review has been seen to take the place of where humans were (2x mandate paper), and many people believe AI can do everything humans can but better. this is a hot debate 

% NOTE: popescu finds higher merge rates for agents vs. humans, pham finds
% higher merge rates for humans
It is highly contested how code review should be adapted to meet these
pressures. 
% \fixme{particularly, how much of it could itself be automated by AI} 
% Practitioners debate how humans should remain involved in code
% review, if at all. 
Reductively, a debate has emerged with two sides: one that views human review
deliberation processes as the bottleneck
preventing software development from
realizing promised productivity gains from AI, and one that believes humans must
remain accountable and embedded in the review process at every step. 
% In February
% 2026, StrongDM AI released press on their Software Factory, a system with
% ``non-interactive development where specs + scenarios drive agents that write
% code, run harnesses, and converge without human review''.
Some researchers claim agents supersede human inspection, declaring ``the end of code review''~\cite{monperrus2026end}, and companies (e.g., OpenAI, StrongDM) have gone further, building agent-driven factories which develop software with little to no human involvement~\cite{openai2026harness, strongdm2026factory}. 
One edict in StrongDM's agent instructions makes the position unmistakable: ``Code \textbf{must not} be reviewed by humans.''
% Could include the "Code must not be reviewed by humans" edict. God I hate
% reading AI-generated blog posts 
Online discussions are mixed, critiquing the
quality of public factory code, asserting humans ``own the merge button'', or arguing for human review of select concerns~\cite{github2025codereview, hnwhyfactoriesfail, osmanicodereviewai}.
On the other side, guidelines from OWASP and Debian \textit{require} human review for all PRs, 
citing security concerns and a need for human responsibility~\cite{owasp2026securecodingai, debian2026vote002}. 
% Some
% researchers seek to advance human-AI alignment and urge benchmarks and agentic
% research to include humans in the loop~\cite{}. 
Some researchers claim AI should be used to \textit{augment} human abilities~\cite{haupt2025position} or recommend mechanisms to force
reviewers to engage critically, as ``oversight degrades the
overseer''~\cite{mitchell2026ai}. Ultimately, no one solution is universally accepted.

Much of this debate is litigated by vendors, standards
bodies, researchers, and anecdotal evidence from social media accounts that may
not extend to real software development, general opinion, or multiple contexts. 
Less is known about challenges AI-assisted authoring brings to reviewers themselves, or which strategies may help.
In this work, we study the reviewer's perspective, as they witness review challenges and organically iterate on promising strategies. We ask:
\begin{itemize}
    \item[\textbf{RQ1:}] How is AI shifting the role of the human reviewer?
    \item[\textbf{RQ2:}] What new challenges do reviewers face from AI-assisted authoring?
    \item[\textbf{RQ3:}] How, and how effectively, is AI used for review?
    \item[\textbf{RQ4:}] Which strategies involving organizational practices, review workflows, and reviewer attention are reported to effectively mitigate review challenges?
\end{itemize}

% i also want the above to talk about challenges why do mixed-methods? i want
% that to be insight #2 also i want it to talk about disruption, not just
% adaptations
To answer these questions, we did a two-phase, mixed-methods study at a large, multinational company with widespread access to AI coding tools.
First, we did semi-structured interviews with developers with
high review load ($n = 24$) to gather contextual experiences of challenges and
strategies.
% and code review observations of a subset of these developers ($n =
% 8$) to supplement self-reports. 
We then conducted an anonymous survey of
randomly-sampled developers ($n = 2{,}590$) to see how the severity of review
challenges and efficacy of review strategies generalize. 
% Anonymity
% allowed us to gain uncensored experiences, while our random sampling recruitment
% allowed us to generalize our findings to a wide range of contexts. 
%
Our study is grounded in the freely-spoken experiences of diverse
practitioners amid this sea change, investigates 
\textit{why} challenges may occur or strategies may work, and demonstrates the magnitude of challenges and strategies at scale. To our knowledge, we contribute the first account of how developers are experiencing code review challenges from AI-assisted authoring and adapting their practices and roles as reviewers in response.
%(RQ 1).
Key findings include:
%
% The two-phase
% design allowed us to triangulate and generalize findings that would not have
% been found by any one method alone. 

% \jenna{is it weird to say "two-phase" and then the work triagulate (which sort of implies 3?)}
% kln: maybe, it seems like people don't like that word so i'll remove it (you are not the first person to dislike it here). it's ok semantically though, used a lot in this context (i think cuz in triangulation you find a third thing -- ``the truth'' -- via the relative positions of two others)

% things qual does: experiences of diverse practitioners, qual allows a
% contextual understanding of why, obs confirms quant: scales, random sample.
% things whole study does: real practitioners who are highly influenced by AI

\begin{enumerate}
\item Most surveyed developers (71\%, vs. 14\% who disagreed, $p < .001$) reported that their \textbf{code review has shifted to focus on higher-level concerns}, such as design and intent. Interviews reveal that AI adoption \textit{enables} this shift by clearing up low-level issues, but also \textit{necessitates} and \textit{complicates} it with increased review burden and shallow review.
\item 
Developers reported that AI-assisted authoring increases review burden via \textbf{PR quality issues}, \textbf{productivity pressures}, \textbf{poor author stewardship}, and \textbf{PR volume and size}. This can lead to fatigue and shallow or skipped review, potentially accumulating debt. 
\item 
Developers reported that AI review aids quality and speed and used it for different purposes, such as generating review comments, understanding PRs, and prioritizing attention. However, in our survey, \textbf{AI review adoption was not correlated with less difficulty keeping up with review} volume and pace, and developers had concerns about overreliance.
\item The strategies reported as most helpful involve \textbf{organizations promoting review and improving its conditions} (e.g., by documenting implicit knowledge for AI tooling) and \textbf{shifting review work upstream to authors and tooling} (e.g., through a pre-review step). Interviewees thought these strategies, along with prioritizing reviewer attention, could free reviewers up to review for important, high-level concerns. 

\end{enumerate}

Based on our findings, we provide specific recommendations for organizations, review workflows, and tooling (Section~\ref{discussion}), such as requiring authors to iterate with AI review prior to PR submission. 

% - ai for code generation has led to increased review volume 
% - no one can agree how the practice of code review should change
% - p3 is a nested point within p2. no one can agree how code review should change and no one can agree on how ai is causing code review challenges 
% - warn about both problems in p1 

% 2) say there are theories of consequences but the full scope is unknown in the first paragraph, punt 

\section{Background and Related Work} 
We present (1) background on code review and (2) related work on AI-assisted authoring, automating review with AI, and reviewers'
perspectives on AI-authored PRs.
\subsection{Code Review}
Code reviews are peer reviews of authored changes to a codebase and related artifacts (e.g., tests, documentation, etc.), with goals such as defect detection, code improvements, knowledge transfer, 
% (an especially important goal in commercial development~\cite{bosu2016process})
increased team awareness, and shared code ownership~\cite{bacchelli2013expectations}. Modern code review evolved as a lightweight alternative to code inspection, an intensive, synchronous process involving structured meetings and defined roles (e.g., moderator, reviewer)~\cite{fagan1999design}. 
% Automation of review tasks has largely targeted technical benefits~\cite{yang2026roadmap}. 
Code review is often integrated with version control tooling and spans from when a PR is opened 
% WRW suggests \emph{pull request} when defining the keyword, even though your audience will know it. 
by an author until it is merged or closed. An author submits a PR, and one or more reviewers are selected by heuristic-based tooling~\cite{ccetin2021review} or by the author. Reviewers inspect the changes (the \textit{changeset}) and leave \textit{review comments} on relevant lines, to which authors respond and make revisions accordingly. Reviewers then decide to reject, accept, or wait for further iteration, until the PR is ultimately accepted, rejected, or closed. Common challenges of code review include review latency and changeset size, and best practices include giving timely feedback and breaking up PRs~\cite{macleod2017code}. In Section~\ref{comparisontopreai}, we compare established challenges, best practices, and outcomes against those found in our work.

\subsection{AI-Driven Changes to Code Review}
AI-assisted authoring has been linked to changes in PR volume, code churn, review latency,
review depth, and code quality~\cite{he2026ai, yu2026habituation, faros2025aiproductivity, li2025rise, appajodu2026scaling, googlecloud_ai_writes_code, he2026speed}.
Research has also identified how humans review and interact with AI-authored PRs (e.g., the degree of required effort)~\cite{khelifi2026behind, duma2026these}. Researchers have built or ideated on LLM-based systems that
generate and incorporate review comments~\cite{tufano2024code,
pirouzkhah2026price, jaoua2025combining, watanabe2024use, ren2025hydra,
li2025codedoctor, tang2024codeagent, lin2025codereviewqa, lin2026leveraging}
or support social aspects of review such as tone and
knowledge transfer~\cite{rahman2024words, heander2025support}. Research mining software repositories 
has found human and AI comments complementary in value offered~\cite{zhong2026human,
goldman2025types} and characterized merge rates of AI-reviewed PRs and incorporation rates of AI
comments~\cite{chowdhury2026industry, sun2026does}. Evidence on the effects of AI on review
quality and speed is mixed~\cite{tufano2025deep, cihan2025automated}.

Closest to our work, grey literature
studies have synthesized a causal theory of AI's team-mediated effects on
review~\cite{agarwal20263100} and practitioners' challenges with code review and expectations for AI review~\cite{davila2024tales}, finding potential effects on code plausibility, review workload, and knowledge transfer. One survey captured how developers believe AI will change code review (e.g., hoping for continued human  oversight)~\cite{dorner2025quo}, and
Alami et al.\ found AI review tools weaken peer review's collective
accountability due to aspects such as absent reciprocity and low human interaction~\cite{alami2025accountability}. A recent literature synthesis surfaced potential sociotechnical risks of AI adoption in code review and 
identified human--AI collaboration in review as an
understudied topic~\cite{yang2026roadmap}.

% Closest to our work, Agarwal et al.\
% synthesized grey literature into a causal theory of AI's team-mediated
% effects on review~\cite{agarwal20263100}, though this research may be acontextual, censored, or biased
% due to online community dynamics~\cite{}. 

Prior work establishes that AI-assisted authoring and reviewing are changing
code review, but practitioner perspectives come largely from potentially
biased or acontextual grey literature~\cite{newman2025disclosure, hu2009overcoming}, and no work investigates strategies practitioners use beyond using AI for review. We address both gaps with a study on reviewer perspectives, identifying the
challenges they face from AI-assisted authoring and the full range of
strategies they use in response.
% These methods also cannot capture certain multidimensional outcomes and
% mechanisms involved in code review (e.g., how much of a PR is AI-authored,
% % aspects of code quality or knowledge transfer, cognitive surrender, etc.).
% Recent work has developed a causal theory for how AI is changing code review
% based on qualitative analysis of grey literature~\cite{}, but this research may be acontextual, censored, or biased
% due to dynamics of online communities (e.g., negativity or social desirability
% biases~\cite{}). 

\section{Methods}
To understand code review challenges from AI-assisted authoring and how review may be effectively adapted in response, we used a two-phase \textit{exploratory sequential} design~\cite{creswell2007mixed}: (1) \textbf{interviews} to qualitatively surface themes and understand the relationships and reasoning behind them, followed by (2) a \textbf{survey} to quantitatively generalize these findings across a large, diverse population of developers. The study was conducted at a large, multinational company that makes varied software products, where developers have widespread access to AI and tokens and encouragement to experiment. Both phases were approved by our company's internal ethics review. 
% and were GDPR-compliant (e.g., honored the right to be forgotten).

% \emerson{For the study setting, replace ``AI-forward'' with concrete characteristics, to the extent accurate and compatible with anonymity: a large multinational firm making varied software products and services, including AI developer tools that are also available internally. Encouragement to experiment and generous access to tokens may help explain why participants build their own agents. We do not need all of this in the introduction, but readers need enough context to interpret the findings.}

\subsection{Phase 1: Interviews}
The first author conducted 50-minute semi-structured interviews
with 24 developers (referred to as P1--24) on Microsoft Teams. 
% P9 also
% participated in a follow-up for his interview. 
To develop the interview protocol, the first author had
informal discussions (pilots, excluded from results) with seven stakeholders (e.g.,
developers, product managers) and one group about AI-driven code review challenges, strategies, and
changes to reviewing practices. 
Interviews progressed through four topics related to participant
contextualization and our RQs: (1) \textit{Background and Review Goals}, 
% (e.g.,
% ``What is the purpose of code review for your team?'')
(2) \textit{Code Review Challenges}, (3)
\textit{Adaptations and AI Review Usage}, and (4) \textit{Wishlist}
(e.g., ``Are there ways review processes or tools could be changed to better support code review?''). Interviewees were asked to compare their experiences to
those before AI-assisted authoring throughout. Our interview protocol is in
our supplementary materials.\footnote{Anonymized supplementary materials: \url{https://doi.org/10.5281/zenodo.24204086}}

% Observation durations were chosen by participants in the range of 25--60
% minutes. Participants completed review responsibilities normally while thinking
% aloud through their process and decision-making. 
% % At the ends of sessions, the study runner interjected to clarify their
% % understanding and probe about the session's ecological validity. \fixme{might
% % be too much info}
% Observations were done to \textit{supplement} interview themes with examples
% from real-life reviewing practices, not to generate new themes or RQs. 

\subheading{Recruitment and Population.} 
% \fixme{how recruited: and in 2 cases, snowball sampling and a pilot who agreed
% to be formally interviewed}
Our goal was to interview developers who experience AI-accelerated development
burdens more acutely and use AI for review in diverse ways.
To reach developers with the
heaviest review loads, we recruited P1--16 and P18 from two large, multi-project repositories on a company DevOps platform, messaging developers in waves of five per repository in descending order of PRs commented on over the prior six weeks.
After 16 interviews, themes had saturated, but some AI usage patterns seen in pilots had not surfaced. To fill this gap, we recruited P17 and P19--24: one via snowball sampling, one pilot participant, and five whose recent review
comments indicated AI assistance. Developers who agreed to participate filled out a consent form and screener on
their review load, AI review usage, demographics, and eligibility. Eligible participants had done code
review professionally before AI-authored code entered PRs they review, so they could compare before and after. 
% All 24
% eligible participants were interviewed. 
% Of the interviewees who were willing to
% be observed, we reached out to and observed 8 whose AI review usage represented
% distinct archetypes identified in interviews (see Section~\ref{} for results on
% review archetypes). 
Participants were men ($n = 17$), women ($n = 5$), or unknown or preferred
not to say ($n = 2$). They were from the United States ($n = 16$), India ($n = 2$), and Norway, Australia, Croatia,
Brazil, Ireland, or unknown (one each). They were aged $26-57$ years (median of $36.5$) and had $3-30+$ years of professional software development experience (median of $14$). 
% Due to nonresponse, attributes beyond nationality
% and job discipline were not collected for P24. 
% Our supplementary materials contain an interview participant table.

\subheading{Analysis.} 
The first author qualitatively analyzed interviews using inductive, \textit{reflexive thematic
analysis}~\cite{braun2006using} in parallel with data collection. They created analytic memos during all interviews and open coded the first two interviews to develop an initial codebook, which they iteratively refined and applied to all 24 interviews using
ATLAS.ti~\cite{atlasti}. The codebook was hierarchical,
with top-level codes organizing material by topic for easy retrieval and
lower-level codes identifying specific concepts (e.g., \texttt{code review for
ai challenges: PR quality}). Codes were split,
reinterpreted, and combined during analysis. 
% For example, \texttt{ai for code review strategies: give context} was
% split into multiple strategies represented in the survey (such as ``Instruct AI
% review to adjust the length, tone, or format of comments'' and ``Give AI review
% access to more relevant code or supporting information''). 
All authors met for an hour weekly to challenge and augment the first author's
interpretations. We did not calculate inter-rater reliability (IRR), consistent
with best practices in HCI when the codebook is part of the analysis process,
rather than its product~\cite{mcdonald2019reliability}. 
% and our method's underlying assumption that
% researcher subjectivity is a resource, rather than a source of bias~\cite{}. 
We avoid counting qualitative items due to our pursuit of unexpected findings
during our inductive analysis, in line with recommended
guidance~\cite{hannah2011counting}; our survey gives quantitative support for themes. 
% Preliminary results from both study phases
% were shared with all participants to elicit feedback on whether results reflected their experiences (i.e.,
% lightweight member checking~\cite{corbin2008basics}). Interview participants reinforced results
% (e.g., noting that they were expected but encouraging to see documented), adding
% confidence that they reflect participants' experiences.

% Analysis of observations was lightweight, as their purpose was to give context
% to interview themes and review tooling. The first author took memos during
% observations and identified timestamps where participants exhibited behavior
% extending an interview theme, later returning to write detailed descriptions for
% segments included in this work.

\subsection{Phase 2: Survey}
\begin{figure*}
    \centering
    \includegraphics[width=\textwidth]{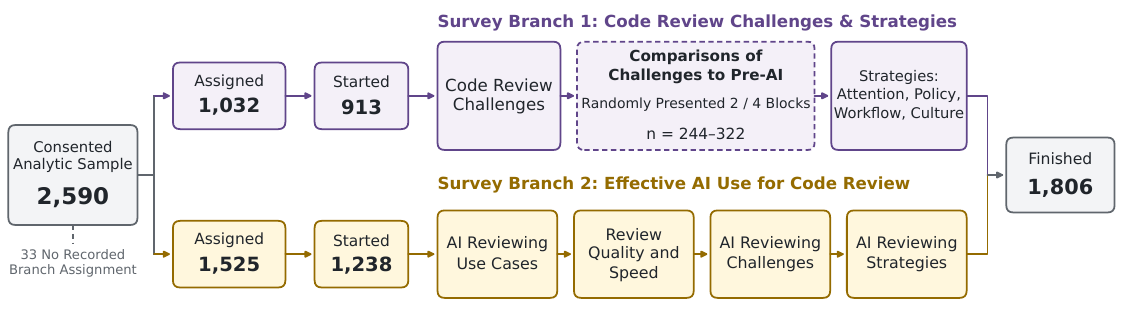}
    \vspace{-30pt}
    \caption{Survey logic flowchart, including participant numbers.}
    \Description{Flowchart of the survey. 2,590 consented respondents split into
    two branches; 33 had no recorded branch assignment. Branch 1, Code Review
    Challenges and Strategies: 1,032 assigned and 913 started; respondents
    answered questions on code review challenges, then two of four randomly
    presented blocks comparing challenges to before AI (244 to 322 respondents
    per block), then strategies for attention, policy, workflow, and culture.
    Branch 2, Effective AI Use for Code Review: 1,525 assigned and 1,238
    started; respondents answered questions on AI reviewing use cases, review
    quality and speed, AI reviewing challenges, and AI reviewing strategies.
    Both branches converge on 1,806 respondents who finished.}
    \label{fig:surveylogic}
\end{figure*}

We designed an anonymous, branching Qualtrics survey to generalize qualitative findings, with items derived from major themes and codes in our codebook. 
Both branches shared consent, eligibility, basic AI
review usage, and demographics questions, but were otherwise self-contained
modules about (1) \textit{Challenges With Code Review and Strategies}
(the ``challenges branch'', designed to take 15 minutes) or (2) \textit{Effectively Using AI for Code
Review} (the ``AI usage branch'', 10 minutes). 
% The challenges branch was designed to take 15 minutes and the AI usage branch, 10 minutes. 
Participants with no AI review experience
were assigned to the challenges branch, while others were randomly assigned
to either. Figure~\ref{fig:surveylogic} shows survey logic and per-branch participant counts. 

Questions were 5-point Likert-type, multi- or single-select, or free response. The challenges branch asked about current code review challenges, how challenges and practices have changed since AI-assisted authoring (``comparison questions''), and strategies for reviewer attention, review workflows, and organizations. Comparison questions were shown only to participants whose professional experience predated AI-assisted authoring. All such participants answered how their ability to keep up with review volume had changed, and then were randomly shown two of four thematic blocks (e.g., PR quality) of comparison questions.
% The remaining comparison questions were grouped into four thematic blocks (e.g., PR quality), with each eligible participant randomly shown two of the four. 
The AI usage branch asked about how AI was used for review, its effects on review quality and speed, its challenges, and strategies to improve its precision and use. Branching and randomized block assignment reduced participant fatigue while preserving statistical power and RQ coverage. Questions were optional except consent, eligibility, basic AI review usage, and
comparison questions.

We iteratively refined both survey branches for clarity, display and branching logic, and duration,
piloting with seven developers, project managers, and researchers recruited
via word-of-mouth and online channels, including
two think-aloud sessions. Pilot data are excluded from results.

\subheading{Recruitment and Sample.} 
We randomly sampled $50{,}000$ users who had posted a comment on any repository on a company DevOps platform in the prior six weeks, then removed known bots and interview participants, to avoid double-counting perspectives. 
% WRW notes: it's probably implied, but it may be worth saying that the "internal" platform
% is at the same AI-forward company as the survey. 
% These were
% downsampled to those that were not on a list of known bot identities.
We recruited via email in waves (first $1\%$ of the sample, $4\%$, and then six of $10\%$) until
response numbers to all questions were sufficiently powerful for planned statistical analyses.
Ultimately, emails were sent to $28{,}130$ users. 
% Interview
% participants were excluded from survey recruitment to avoid double-counting perspectives. 
Survey data collection took place from July 29 through
August 21, 2026. 

We excluded responses with implausible demographics: those older than three standard deviations above the mean, or with years of experience exceeding age minus 14.
% We omitted responses where reported age exceeded 
% the mean age plus three standard deviations, or where professional years of experience exceeded 
% their age minus 14. 
The remaining $2{,}590$ participants who filled out the consent, eligibility, and AI
usage questions form the \textit{analytic pool} of responses represented in this work. $1{,}806$ finished the survey (i.e.,
clicked to the end), resulting in a response rate of $9\%$ for all responses used
and $6\%$ without partial responses. 
Of the $2{,}557$ participants who were assigned a branch, $40\%$ were assigned the challenges branch; the rest were assigned the other.
% The
% challenges branch took participants on average \fixme{X} minutes longer than the
% AI usage branch and \fixme{num$\times$} more participants finished the AI usage
% branch than did the challenges branch. 
Because of branching, optional questions, and the use of partial responses, all
statistics are computed over respondents that answered the
relevant questions. 

\subheading{Population.} Demographic questions were optional, so counts do not sum to
$2{,}590$. Participants were men ($n = 1{,}344$, $81\%$), women
($n = 259$, $16\%$), nonbinary or self-described ($n = 11$, $1\%$), or
preferred not to say ($n = 48$, $3\%$). 
% Unfortunately, the skew toward men reflects the demographics of the computing discipline.
% We discuss this
% limitation and future work on gender bias in code review in Section~\ref{}.
They had $0-45$ years of professional experience (median
of $10$), were aged $20-69$ (median of $34$), and hailed from $45$
countries across six continents: North America ($n=1{,}044$, $64\%$), Asia ($n = 292$, $18\%$), Europe ($n=233$, $14\%$), South America ($n=27$, $2\%$), Oceania ($n = 26$, $2\%$), and Africa ($n = 17$, $1\%$). Because we sampled by code review activity, developers spanned many job disciplines, most commonly ``Software Engineer'' ($n
= 1{,}548$, $90\%$) and ``Software Architect'' ($n = 38$, $2\%$). The median participant reported reviewing $6-10$ hours per week. 

\subheading{Analysis.} We used one-sample Wilcoxon signed-rank tests against the neutral response
to detect directional skew in comparison questions~\cite{hollander2013nonparametric}.
To rank strategies by helpfulness when respondents chose from a subset of those they had used, we fit extended Plackett--Luce models that accommodate sub-rankings and ties~\cite{plackettluce}. All $p$-values shown are Benjamini--Hochberg-adjusted for multiple comparisons within hypotheses and related tests~\cite{benjamini1995controlling}. AI was used to help write analysis scripts.

\section{Results}
We present our results, organized by research question.
% \iffalse

% RQ1
% challenges
% contexts                                  outcomes
% A --\                                  /--X
% B --> (too much burden on reviewer) --+--Y
% C --/                                  \--Z 

% AI writes a lot of --> high reviewing burden --> people don't review & get mad

% RQ2. 
% people THINK they can't do review any more without AI 
% survey: it makes review BETTER but has solved the bottleneck

% use cases (use AI for review, organizations, workflows, ...) 
% WRW "the default star-diagram with too-much-burden" in center and radiating
% out to strategy names + sec #s is NOT needed.

% If that diagram includes "numbers" on the edges, or if conceptually contains information like a pie chart (although reviewers hate pie charts), then it becomes optional.

% ultimately: diagram not necessary

% RQ3. code review strategies
% org practices support CR
% preparing PRs pre-reviewer
% reviewer attentional strategies
% AI tooling to help reviewers

% "things done BEFORE the reviewing" (org practices, preparing)
% "things done DURING the reviewing" (attention, AI tooling)

% * use "AI-Authoring" and "AI-ReviewING" instead of x4y

% KAIA WONDERS: if I move up the stuff from RQ4, do I then have a giant RQ2, or do I split off a little RQ2 bud? 

%     \fi 
%     then it should be clear downstream debt/fatigue are explicit outcomes}}

\noindent\textbf{$\star$ Survey notation:} Survey results on review challenge frequency, and changes since 
AI-assisted authoring, appear inline for Sections~\ref{quovadis} and~\ref{r4ai}. We use the notation (\textbf{ID}, 
[$x$\%,] $y\%/z\%$, \textit{direction}): \textbf{ID} labels the survey item, $x$\% (where 
available) is the percent of respondents who experience a challenge at least half the time, $y\%/z\%$ 
gives the split of non-neutral responses (\textit{more}/\textit{less} or \textit{agree}/\textit{disagree}), 
and \textit{direction} is the significant
direction from the Wilcoxon test ($p_{\text{adj}} < .05$), or
\textit{inconclusive} otherwise. Trends (i.e., $.05\leq p_{adj}\leq.1$) are reported as \textit{weakly} in a 
direction. 

\subsection{RQ1 - The Role of Human Review}
\label{quovadis}
Our results demonstrate a \textbf{shift in the role of the human reviewer toward 
higher-level concerns}, a shift that AI both \textit{enables} and 
\textit{necessitates}, while also \textit{complicating} reviewers' ability 
to make it. Table~\ref{tab:high-level} presents survey data on this shift.
Below, we describe how AI \textit{enables} high-level review. In 
Section~\ref{aireview challenges}, we demonstrate how AI-assisted authoring \textit{necessitates} higher-level review, and how it \textit{complicates} 
quality review by creating increased burden for the human reviewer.

\begin{table}[h]
\centering
\caption{\textbf{Perceived review changes after AI-assisted authoring.} Percentages are respondents who agreed (purple) vs. disagreed (yellow).
$n=322$. Survey items are cut for space.}
\label{tab:high-level}
\begingroup
\fontsize{6.6}{7.6}\selectfont
\setlength{\tabcolsep}{0pt}
\setlength{\aboverulesep}{0.1ex}
\setlength{\belowrulesep}{0.1ex}
\setlength{\abovetopsep}{0pt}
\setlength{\belowbottomsep}{0pt}
\renewcommand{\arraystretch}{1.10}
\newcommand{\KaiaBipolarBar}[7]{%
  \begingroup%
  \def\KaiaColorOne{KaiaCompFive}%
  \def\KaiaColorTwo{KaiaCompFour}%
  \def\KaiaColorThree{KaiaCompThree}%
  \def\KaiaColorFour{KaiaCompTwo}%
  \def\KaiaColorFive{KaiaCompOne}%
  \KaiaDistributionBar{#1}{#2}{#3}{#4}{#5}{#6}{#7}%
  \endgroup%
}
\sbox{\KaiaTableBox}{%
\begin{tabular}{@{}>{\raggedright\arraybackslash}p{0.05\columnwidth}@{\hspace{4pt}}>{\raggedright\arraybackslash}p{0.53\columnwidth}@{\hspace{4pt}}>{\raggedright\arraybackslash}p{0.29\columnwidth}@{\hspace{4pt}}>{\raggedleft\arraybackslash}p{\dimexpr\columnwidth-0.05\columnwidth-0.53\columnwidth-0.29\columnwidth-12pt\relax}@{}}
\toprule
\textbf{ID} & \textbf{Change} & \multicolumn{1}{@{\hspace{4pt}}p{0.29\columnwidth}@{}}{\KaiaDistributionHeader{Distribution}} & \textbf{$p_{adj}$} \\
\midrule
RC1 & My review has shifted toward higher-level concerns (e.g., design and intent). & \KaiaBipolarBar{0.0465838509}{0.0962732919}{0.1490683230}{0.4440993789}{0.2639751553}{14}{71} & $p < .001$ \\
RC2 & I spend more effort verifying that PRs address the right problem. & \KaiaBipolarBar{0.0434782609}{0.1055900621}{0.2919254658}{0.3850931677}{0.1739130435}{15}{56} & $p < .001$ \\
RC3 & Code review is now less valuable for learning new coding approaches. & \KaiaBipolarBar{0.1614906832}{0.1211180124}{0.2173913043}{0.3447204969}{0.1552795031}{28}{50} & $p = .03$ \\
RC4 & PR authors discuss design with me less in advance than they used to. & \KaiaBipolarBar{0.0745341615}{0.1273291925}{0.3571428571}{0.2484472050}{0.1925465839}{20}{44} & $p < .001$ \\
RC5 & PR authors take my comments about their code less personally. & \KaiaBipolarBar{0.0527950311}{0.0652173913}{0.5714285714}{0.2111801242}{0.0993788820}{12}{31} & $p < .001$ \\
\bottomrule
\end{tabular}%
}
\usebox{\KaiaTableBox}
\par\vspace{0pt}
\begin{minipage}{\wd\KaiaTableBox}
\fontsize{6.2}{7.2}\selectfont
\KaiaLegend{KaiaCompFive}{Strongly disagree}\quad\KaiaLegend{KaiaCompFour}{Somewhat disagree}\quad\KaiaLegend{KaiaCompThree}{Neither}\quad\KaiaLegend{KaiaCompTwo}{Somewhat agree}\quad\KaiaLegend{KaiaCompOne}{Strongly agree}\\[-4pt]
\noindent\rule{\linewidth}{\heavyrulewidth}
\end{minipage}
\endgroup
\end{table}

Surveyed developers strongly indicated that their review is transitioning to focus more on higher-level concerns, such
as intent and design (\textbf{RC1}, $71\%/14\%$, \textit{agree}). Interviews revealed how this shift was or will be \textbf{enabled} by the capabilities of AI to eliminate low-level issues in code, freeing the human reviewer to look at concerns AI is less capable of reviewing, e.g., \qualquote{As the lower-level review gets automated and out of the way, it enables for much richer higher-level review, which can bring even more value to the table than it can today} (P9). Interviewees who could trust AI review to clear up many defects or other quality issues (e.g., P1)
% or who authored code in situations where these issues were deprioritized (e.g., rapid prototyping), 
reported reviewing at a high level with less concern. 
% \emerson{``Navigated this shift more effectively'' feels stronger than the evidence presented here. What supports the comparison? Could we show that evidence or qualify this as participants' perceptions?}

Design review may also now be deferred to code review: surveyed developers thought authors \textit{discuss the design of PRs less in advance} of PR submission than before AI-assisted authoring (\textbf{RC4}, $56\%/15\%$, \textit{agree}). Interviews revealed design was not discussed in advance because AI can be used to rapidly prototype and try out designs.
Some interviewees described how their team would review design in tandem with code: \qualquote{Design and code writing can be all done in one single PR\ldots I'm trying to change the perspective
of my team that your baby is the [design] document, not the code itself\ldots that's one way we are trying to reduce the friction and make this cycle shorter} (P22).

Interviewees (e.g., P5) noted that they felt freer to nitpick or skip tone calibration when reviewing, since authors often interact with PRs via agents. Surveyed developers reported that \textit{authors take review feedback less personally} (\textbf{RC5}, $31\%/12\%$, \textit{agree}). Interviewees pointed to two possible explanations: a diminished sense of ownership over AI-authored code (seen in prior work~\cite{martin2026more, seo2026whose}) and, when the reviewer was also using AI, the sense that critique was not coming from a human: \qualquote{You can say, `yes, it's the AI that's complaining about your code, it's not me'\ldots 
the [author] can also say, `I don't know why the AI did that'} (P21). 
% I've got one person whose code I review a lot and I know I'm not going to have to burn political capital or get in a fight with him\ldots I can just let the AI fire the comments in, and he just uses the AI right back} (P21). 
Developers may also no longer see review as a useful channel for learning low-level programming techniques (\textbf{RC3}, $50\%/28\%$, \textit{agree}). 
% though some interviewees (e.g., P19) thought exposure to AI-authored code provides learning benefits.
% However, others (e.g., P23) noted only receiving bot-generated comments can leave repository newcomers feeling unwelcome.

However, this shift in the reviewer's role is \textit{necessitated} and \textit{complicated} by AI-assisted authoring.
% For example, developers reported prioritizing higher-level concerns due to strain on their attention from PR volume driven by AI authoring (see Section~\ref{}).
% \qualquote{Before AI, [review]
% was like, make sure you approve that this is good and there's nothing you would
% have done better\ldots With GenAI, the volume is just so huge\ldots It's a
% delicate balance of not stalling everything, but at the same time catching the
% thing that is not going to hurt other teams} (P9). 
For example, developers spend \emph{more
effort verifying PRs solve the right problem} (\textbf{RC2}, $56\%/15\%$, \emph{agree}) than before, but this may in part be due to \textit{uncertain intent} in PRs (see Section~\ref{r4aichal:stewardship}).

\summarytcolorbox{Reviewers indicated their \textbf{role has shifted toward higher-level concerns} (e.g., intent and design). This shift is enabled, necessitated, and complicated by AI adoption.}

\subsection{RQ2 - Challenges From AI-Assisted Authoring}
\label{r4ai}
% \begin{figure}[h]
%     \centering
%     \includegraphics[width=\textwidth]{figures/Code Review Framework.pdf}
%     \caption{\fixme{Graphic design is not my passion. }\jenna{maybe we can make the figure literally asymmetric to visualize it better?}}
%     % Right = consequences
%     % Left = factors 

%     %           /----> Fatigue
%     % [ ARB] ---+                       
%     %           \----> Shallow & Missed --> Debt

%     \label{fig:r4aichal}
% \end{figure}

Our results show that AI authoring brings many challenges that necessitate shallower review or the focus on high-level concerns. Factors such as a lack of author stewardship and PR volume can lead to an \textbf{increased burden} on the reviewer, leading to fatigue, shallow or missed review, and \textbf{downstream
debt}. Table~\ref{tab:challenges-combined} shows most survey results for Sections~\ref{r4aichal:productivitypressures}--~\ref{r4aichal:attentionstrain}.

\begin{table}[h]
\centering
\caption{\textbf{Code review challenges.} Numbers on the right of the distribution indicate \% ``About half the time'' or more; those to the left indicate \% ``Sometimes'' or ``Never''. Per-item $n=638-905$. More vs. Less ($p_{adj}$) reports changes since AI-assisted authoring (per-item $n=242-317$). Survey items are cut for space.}
\label{tab:challenges-combined}
\begingroup
\fontsize{6.6}{7.6}\selectfont
\setlength{\tabcolsep}{0pt}
\setlength{\aboverulesep}{0.1ex}
\setlength{\belowrulesep}{0.1ex}
\setlength{\abovetopsep}{0pt}
\setlength{\belowbottomsep}{0pt}
\renewcommand{\arraystretch}{0.98}
\sbox{\KaiaTableBox}{%
\begin{tabular}{@{}>{\raggedright\arraybackslash}p{0.05\columnwidth}@{\hspace{6pt}}>{\raggedright\arraybackslash}p{0.38\columnwidth}@{\hspace{6pt}}>{\raggedright\arraybackslash}p{0.27\columnwidth}@{\hspace{6pt}}>{\raggedleft\arraybackslash}p{\dimexpr\columnwidth-0.05\columnwidth-0.38\columnwidth-0.27\columnwidth-18pt\relax}@{}}
\toprule
\textbf{ID} & \textbf{Challenge} & \multicolumn{1}{@{\hspace{6pt}}p{0.\columnwidth}@{}}{\KaiaDistributionHeader{Distribution}} & \textbf{More vs. Less} ($p_{adj}$) \\[0pt]
\midrule
C1 & A PR I reviewed had a changeset that was too large & \KaiaFrequencyBar{0.0740331492}{0.4552486188}{0.2596685083}{0.1922651934}{0.0187845304}{53}{47} & \textbf{60\%} vs. 13\% ($p < .001$) \\
C2 & I felt pressure to review faster & \KaiaFrequencyBar{0.2931415929}{0.2942477876}{0.1537610619}{0.1780973451}{0.0807522124}{59}{41} & \textbf{42\%} vs. 13\% ($p < .001$) \\
C3 & PR lacked evidence that it was adequately tested & \KaiaFrequencyBar{0.1406025825}{0.4634146341}{0.2195121951}{0.1449067432}{0.0315638451}{60}{40} & 34\% vs. 34\% ($p = .85$) \\
C4 & I didn't do as thorough of a review as I wanted & \KaiaFrequencyBar{0.2308546060}{0.4162042175}{0.1920088790}{0.1231964484}{0.0377358491}{65}{35} & \textbf{47\%} vs. 19\% ($p < .001$) \\
C5 & I spent extra time understanding poorly-explained PRs & \KaiaFrequencyBar{0.2030534351}{0.4595419847}{0.1801526718}{0.1328244275}{0.0244274809}{66}{34} & \textbf{42\%} vs. 25\% ($p = .03$) \\
C6 & Code did not follow architectural or stylistic conventions & \KaiaFrequencyBar{0.1954022989}{0.4712643678}{0.1853448276}{0.1120689655}{0.0359195402}{67}{33} & 33\% vs. \textbf{34\%} ($p = .29$) \\
C7 & PR description did not give needed context & \KaiaFrequencyBar{0.2682926829}{0.4045911047}{0.1750358680}{0.1133428981}{0.0387374462}{67}{33} & 20\% vs. \textbf{47\%} ($p < .001$) \\
C8 & I found problems the author should have caught & \KaiaFrequencyBar{0.1780821918}{0.5144596651}{0.1735159817}{0.1050228311}{0.0289193303}{69}{31} & \textbf{35\%} vs. 28\% ($p = .50$) \\
C9 & Review affected my other job responsibilities & \KaiaFrequencyBar{0.3392461197}{0.3780487805}{0.1452328160}{0.0975609756}{0.0399113082}{72}{28} & \textbf{32\%} vs. 21\% ($p = .10$) \\
C10 & Code that was subtly incorrect & \KaiaFrequencyBar{0.1479885057}{0.5804597701}{0.1910919540}{0.0689655172}{0.0114942529}{73}{27} & \textbf{40\%} vs. 26\% ($p = .02$) \\
C11 & Code reimplementing existing functionality & \KaiaFrequencyBar{0.2474820144}{0.4820143885}{0.1769784173}{0.0748201439}{0.0187050360}{73}{27} & \textbf{39\%} vs. 26\% ($p = .09$) \\
C12 & A PR addressed something unimportant or unnecessary & \KaiaFrequencyBar{0.2900432900}{0.4473304473}{0.1587301587}{0.0880230880}{0.0158730159}{74}{26} & \textbf{38\%} vs. 25\% ($p = .09$) \\
C13 & Lack of author reasoning needed to review effectively & \KaiaFrequencyBar{0.2583586626}{0.4954407295}{0.1367781155}{0.0835866261}{0.0258358663}{75}{25} & 31\% vs. \textbf{33\%} ($p = .24$) \\
C14 & The author could not explain or justify a change & \KaiaFrequencyBar{0.3605015674}{0.3949843260}{0.1473354232}{0.0862068966}{0.0109717868}{76}{24} & \textbf{49\%} vs. 26\% ($p = .03$) \\
C15 & The same kinds of mistakes appeared across PRs & \KaiaFrequencyBar{0.3366906475}{0.4374100719}{0.1136690647}{0.0877697842}{0.0244604317}{77}{23} & 24\% vs. \textbf{32\%} ($p = .03$) \\
C16 & Cultural friction because of differing review expectations & \KaiaFrequencyBar{0.4390625000}{0.3453125000}{0.1156250000}{0.0750000000}{0.0250000000}{78}{22} & \textbf{25\%} vs. 23\% ($p = .47$) \\
C17 & A PR I wanted to review did not need it when I looked & \KaiaFrequencyBar{0.2214839424}{0.5692137320}{0.1583610188}{0.0476190476}{0.0033222591}{79}{21} & 13\% vs. \textbf{14\%} ($p = .28$) \\
C18 & I needed to repeat feedback I'd given on other PRs & \KaiaFrequencyBar{0.3955555556}{0.4255555556}{0.1011111111}{0.0655555556}{0.0122222222}{82}{18} & \textbf{23\%} vs. 21\% ($p = .80$) \\
C19 & A PR I wanted to review was merged without my review & \KaiaFrequencyBar{0.3270509978}{0.5199556541}{0.1053215078}{0.0388026608}{0.0088691796}{85}{15} & \textbf{19\%} vs. 14\% ($p = .73$) \\
C20 & The author's changes did not address my comments & \KaiaFrequencyBar{0.4673748103}{0.3930197269}{0.0773899848}{0.0470409712}{0.0151745068}{86}{14} & 24\% vs. \textbf{31\%} ($p = .03$) \\
C21 & I felt that the author was not being respectful of my time & \KaiaFrequencyBar{0.6294573643}{0.2372093023}{0.0697674419}{0.0527131783}{0.0108527132}{87}{13} & 22\% vs. \textbf{31\%} ($p = .04$) \\
\bottomrule
\end{tabular}%
}
\usebox{\KaiaTableBox}
\par\vspace{0pt}
\begin{minipage}{\wd\KaiaTableBox}
\fontsize{6}{7}\selectfont
\KaiaLegend{KaiaFreqOne}{Never}\quad\KaiaLegend{KaiaFreqTwo}{Sometimes}\quad\KaiaLegend{KaiaFreqThree}{About half the time}\quad\KaiaLegend{KaiaFreqFour}{Most of the time}\quad\KaiaLegend{KaiaFreqFive}{Always}\\[-4pt]
\noindent\rule{\linewidth}{\heavyrulewidth}
\end{minipage}
\endgroup
\end{table}

\subsubsection{Factors: PR Quality and Productivity Pressures}
\label{r4aichal:productivitypressures}
We detail upstream factors contributing to increased burden on reviewers,
including \textbf{PR quality issues} and \textbf{productivity pressures}.

\subheading{PR Quality Issues.} Developers noted that AI 
% \jenna{AI review? because you mention it only knows what is in a PR which is only the case in review, not generation. If not maybe something like ``Developers noted that AI can be imprecise due to \emph{lack of situation or codebase awarness} at review time, and lack of \emph{out of code base knowledge} (such as tribal knowledge contained in teams) during generation"} 
% kln: i'm struggling with this because i heard the same thing about both ai authoring and review, but it's brought up later in the rq3 results, so i'll just display it in terms of authoring here maybe??
was imprecise and traced this back to its \emph{lack of situational or
codebase awareness} (e.g., lack of knowledge contained in teams). 
Interviewees noted AI authoring or reviewing misses important context more often in \emph{legacy codebases} that may have
uneven architecture or use old language patterns or API calls, e.g.,
\qualquote{
% What [AI is] not good at is knowing what to build\ldots An easy example is
[In] codebases that are brownfield, where there’s some new stuff, some old stuff, you
want to match the new patterns, not the old patterns. That’s not always easy
for AI to tell} (P1). Similarly, this problem can arise in \emph{large codebases}: \qualquote{Because
I work in humongous repos, we're talking about 22,000 to 25,000
projects in a repository, AI\ldots 
% may say, `well, this is wrong'\ldots but
% that\ldots may [not] be quite correct for a specific
% repository. It 
may not be able to go and find, hey, in this project we've
done it this way} (P23). 
AI may also exhibit more precision issues when rare patterns or conventions are common
and desired in a codebase (e.g., Hungarian variable naming, for P5).

% \jenna{There is a decent amount of repetition still in the results.. maybe when you're done we should give it all to a model and ask it to show areas where we can bring together?}\\
% kln: agree, you can do this or determine which things are most important at any time
Developers reported a rise in PRs that are \textit{subtly incorrect} (\textbf{C10}, $27\%$, $40\%/26\%$, \emph{more}), which could be amplified with authors who
lack relevant expertise with the codebase (e.g., junior engineers or cross-team partners, according to P18). Interviewees noted these PRs may have signals of
correctness (e.g., passing tests or builds) or
fix a local issue,
but require domain expertise to understand negative downstream impacts.
P12 described one such \qualquote{insidious} PR: \qualquote{[AI] fixed the problem\ldots it actually would have introduced another regression how everyone
views mail like 99\% of the time\ldots We only realized that after it got checked in.}
% it looks legit and actually
% doesn't violate anything at all in the code and is the right way to do it.
Developers also reported AI-driven challenges with  \textit{missed code reuse opportunities} (\textbf{C11}, $27\%$, $39\%/26\%$, \textit{weakly more}). 
% as \qualquote{agents still tend to write new
% things
% versus reusing code} (P1). 

Developers also reported code
non-conformant with desired or conventional style and architecture, e.g., one PR
where P5 could
\qualquote{clearly tell that\ldots it was entirely
vibe-coded\ldots this just isn't how we do things
in [our product]\ldots} 
However, survey results did not show a significant change in \textit{unconventional code} (\textbf{C6}, $33\%$, $33\%/34\%$, \textit{inconclusive}). 
% \jenna{If we are worried about space, this paragraph doesnt have a super interesting finding}
% \fixme{kln: heard, we can cut}
% Interviewees (e.g., P5)
% also noted that AI-generated PR descriptions may describe what was changed, rather than why. By contrast, surveyed developers reported
% PRs with \textit{inadequate description or title context} had decreased in frequency (\textbf{C7}, $33\%$, $20\%$/$47\%$, \textit{less}). 
% Interviews revealed
% that AI may raise the existing floor: \qualquote{Before AI\ldots 
% people hardly wrote anything.
% But now people write a lot in descriptions. Sometimes not very useful stuff, but
% at least they do write stuff} (P18). 
% \fixme{P5 had autogenerated
% descriptions, but they turned them off because it was making people not put
% effort into the descriptions/made them worse overall}
Interviewees (e.g., P10) also noted that AI-generated artifacts \emph{lag with iteration}, which 
can lead to extra work. 
% These new challenges may not generalize due to some interviewees reviewing in
% contexts where 
% low-precision AI authoring is more prevalent (see Section~\ref{}) or, as some
% interviewees noted (e.g., \fixme{pid}), AI authoring can raise the number
% of PRs submitted that have tests at all. 

\subheading{Productivity Pressures.} Surveyed developers felt pressure to review faster than before (\textbf{C2}, $41\%$, $42\%/13\%$, \textit{more}).
Interviews revealed this can be motivated by review volume, wanting to unblock others
(e.g., given increased review latency), team culture, or pressure from authors, e.g.,
\qualquote{People are in a rush---they want an approval, they don't want a
review} (P9). Productivity pressures also stem from perceived organizational
pressures to increase throughput based on AI capabilities: \qualquote{Leaders just assume that AI can
do something in 40 minutes. It's just not true. Yeah, I might be able to pump
out code, but code reviewing\ldots that's not factored into anything} (P24).
Interviewees also mentioned 
% that AI has \qualquote{creat[ed] an expectation from
% the top down that speed is more important than correctness} (P4) and 
that
quantitative metrics, such as the need \qualquote{to merge 3 PRs a week} (P16),
may be more valued on performance reviews. Interviewees indicated these pressures could lead to
shallower or missed review and downstream debt (see Section~\ref{r4aichal:attentionstrain}).

\subsubsection{Factors: Lack of Stewardship and Uncertain Intent} 
\label{r4aichal:stewardship}
We found that AI authoring can lead to a \textbf{lack of author stewardship} and \textbf{uncertain intent}, creating burden for the reviewer. 

\subheading{Lack of Author Stewardship.} Developers reported that AI allows authors
to submit PRs they do not understand, resulting in authors being
\emph{unable to answer questions} about their PRs (\textbf{C14}, $24\%$, $49\%/26\%$, \textit{more}) 
and reviewers needing to spend \textit{additional effort understanding changes} due to inadequate explanation from the author (\textbf{C5}, $34\%$, $42\%/25\%$, \textit{more}).
\qualquote{My biggest challenge in reviewing\ldots people these days, they don't look at the
code they're producing...you ask [the author], oh, ‘why didn't you do this?’ Oh, ‘AI
just did this and I left it like that.’ That makes me a little sad} (P15).
Interviewees (e.g., P4) noted this could result from overtrust in, or \textit{cognitive surrender}~\cite{shaw2026thinking} to, AI.
% : 
% \qualquote{they're trusting AI, and AI doesn't know} (P4). 
% \jenna{The below paragraph seems to just contradict itself without explaining why}\\

Interviewees also experienced difficulty with authors incorporating their
comments improperly or being intermediaries for their authoring agents, e.g., \qualquote{I was trying to
coach the AI into doing something and then at the end of the day, it wasn't able
to do it properly\ldots It’s \textbf{more responsibility on the people who are
reviewing and zero responsibility on the authors.} And that shift in
responsibility has led to a lot more of my time being in these PRs\ldots 
% It's
% very common to get to 35, 36, 50 iterations on a code review
} (P6). Interviewees (e.g., P14) noted seeing issues (e.g., large changesets) that authors could have caught, expressing frustration. Interviewees noted
that they see \emph{repetitive mistakes} across PRs more often 
due to the 
difficulty of getting authors and their agents to learn
from reviewer feedback, e.g., 
\qualquote{I block a PR and then a month later I get
another very similar PR 
% by a different engineer 
% and I need to go block it again
\ldots
[AI] doesn't have the context\ldots 
% and the
% history from the previous PRs and the other engineers
} (P2). 

By contrast, surveyed developers reported experiencing \textit{authors failing to incorporate comments}, \textit{a felt lack of respect from authors}, and seeing \textit{repetitive mistakes} less often than before AI-assisted authoring, and infrequently (e.g., \textbf{C20}, $14\%$, $24\%/31\%$, \textit{less}, as well as \textbf{C21}, \textbf{C15}, and \textbf{C18}). Surveyed developers also did not report significant changes in \textit{preventable PR quality issues} left by authors (\textbf{C8}) or \textit{cultural friction} based on the expected rigor of code review (\textbf{C16}). These divergences may relate to differing attributes between interview and survey samples (discussed in Section~\ref{limitations}).

\subheading{Uncertain Intent.} Developers reported that AI-generated PRs often have
\textit{misaligned or uncertain intent} (i.e., they lack justification or seem
useless), e.g., \qualquote{LLMs love to do like, `you won't need it', kinds of things\ldots when you
review, you don't really know whether this was the intent of the author. I
find myself commenting, `do we mean to do that?' And often the answer is
no\ldots this is [an] introduced maintenance cost} (P9).
% This is this is like not applicable
% for our scenario. And that's true also for the code that is produced, for the
% automatic PR review comments, that also pinpoint some issues that really don't
% apply at all in the way we use that tool. So this is a waste of time to add
% complexity around this. So Yeah, the to find the, you know, to add a lot of
% things that is not tested, that is never, you know, is that code, but that we
% need to maintain. But that adds, you know, a burden on the full system, so
% that's uh. .. Yeah, it's a bit hard to fight that.
Interviewees (e.g., P15) suggested they spend more time now reconstructing
intent, 
% e.g., \qualquote{[I'm] spending much more time on trying to do the reverse engineering of what is
% being done. I think a big part of the code review process is
% putting yourself in the person's shoes 
% % and 
% % \ldots understanding why they're doing that
% % change
% \ldots when it's AI,
% it's probably whatever AI reasoned about in that moment, then it doesn't always
% make sense} (P15). 
and survey results agree, suggesting that developers more often encounter
\textit{seemingly-unimportant PRs} (\textbf{C12}, $26\%$, $38\%/25\%$ \emph{weakly more}). However, they may not be less \textit{able to review effectively due to lost intent} (\textbf{C13}, $25\%$, $31\%/33\%$, \textit{inconclusive}).
% P12 reflected
% on a meeting where he had met with leaders to discuss AI policy:
% \qualquote{It sounded like the
% largest source of review fatigue for a lot of the people in our org is
% that now instead of just reviewing the code to be like, `are you doing this in
% the right way architecturally?' You also have to review, `is this even the right
% change at all?'} (P12)

Interviewees reported that lacking knowledge about the author's AI usage
can lead to \textit{difficulty calibrating review scope and depth}. Interviewees noted this calibration is often based on social factors related
to risk, such as the author's historical PR quality, seniority, and expertise. However, these heuristics can be useless
when the degree of
author engagement is unknown: \qualquote{In the past, if you knew\ldots that [somebody] typically
write[s] code that works\ldots you don't review it as
rigorously\ldots now it's like every PR needs to have the exact same scrutiny as if
an intern has written the code} (P8). 

\subsubsection{Factors: Within- and Between-PR Attention Strain} 
\label{r4aichal:attentionstrain}
Developers reported that AI-driven PR size and volume can lead to \textbf{attention strain}, leading to shallow or missed review. 

\subheading{Within-PR Attention Strain.} Developers noted that AI-assisted authoring has increasingly led to \textit{large} PRs 
(\textbf{C1}, $47\%$, $60\%/13\%$, \emph{more}), making each PR harder or impossible to parse, e.g.,
\qualquote{One of the problems I have is the sheer size of some of the
reviews\ldots you have to build all this context around what's changing, and it
can be really high friction\ldots} (P11). 
Given PR volume and how review scope has increased,
some interviewees noted \qualquote{context switching is higher} (P10), though others (e.g.,
P2) noted it
was an expected part of their role or a personal strength. 
% chunk prs/gate prs of a certain size juliana: more code == more liability
Surveyed developers
indicated
they worry their \emph{review is shallow} (\textbf{C4}, $35\%$, $47\%/19\%$, \emph{more}), and interviewees
commented that \qualquote{lots of people just rubber stamp PRs} (P13) (i.e.,
approve PRs as a formality). 
% \jenna{but this wasn't backed up in the survey right? So maybe we just had grumpy/worried people in interviews? }
% kln: this particular thing was backed up in the survey - people were worried their review is shallow more now than before

\subheading{Between-PR Attention Strain.} 
All interviewees mentioned challenges keeping up with the \emph{volume and pace} of
code reviews, a difficulty strongly corroborated by our survey: $3.4\times$ more respondents agreed than disagreed
that keeping up had been difficult. Surveyed developers reported this difficulty has worsened ($56\%/22\%$, \textit{more difficult}).

Interviewees mentioned that PR volume and
engineering system friction
have increased
\emph{review latency}, e.g., 
\qualquote{I could produce five PRs a day\ldots but now I'm going
to be paying\ldots ten times the cost of\ldots handling code review feedback,
re-queuing my builds, babysitting the pull requests\ldots it holds me back} (P14). 
Interviewees mentioned
frustration with \textit{review routing} exacerbating this latency (\textbf{C17}, 21\%, $13\%/14\%$, \emph{inconclusive}).
Interviewees (e.g., P17) indicated
this is an old challenge, but that PR volume
has made it more pressing.
% , 
% e.g., \qualquote{I think the
% difficulty level is the same, but the problem is aggravated by the sheer volume
% of pull requests\ldots 
% I don't want to have to figure out which channel to join, who to
% DM\ldots [it] actually takes a lot of engineering time} (P17). 
Authors may also send PRs out before they are ready, causing \textit{notification noise}:
\qualquote{People are more likely\ldots to push out what the agent said, and then
ping people for code reviews\ldots and
then [expect] other people to be the checks and balances} (P7). 
This noise was noted to be
increased by review latency (e.g., by P9) and productivity pressures (e.g., by P24).
% AI's attempt at cutting:
% Interviewees noted that authors may submit PRs earlier and remind reviewers more frequently to try to merge their PRs, contributing to [notification fatigue? inefficient review routing?] as authors iterate or CI builds fail. As P7 put it, people increasingly \qualquote{push out what the agent said} and ping for review \qualquote{way sooner than they usually would}, expecting others to serve as \qualquote{checks and balances}. This was also linked to increased review latency (e.g., P9) and productivity pressures (e.g., P24).

Developers noted attention strain and productivity pressures 
can lead to PRs \emph{slipping
through} without desired human review (\textbf{C19}, $15\%$, $19\%/14\%$, \emph{inconclusive}), e.g., \qualquote{I do a lot more reviews now because they come in very,
very quickly and people are expected to ramp up on code almost
instantly, so things are getting pushed that aren't ready} (P6). Interviewees (e.g., P2) noted this is something they
usually recognize in hindsight. 
% the logic behind having this last is that when overflowing reviewer capacity,
% slippage occurs and downstream issues

\subsubsection{Consequences: Fatigue and Debt}
\label{r4aichal:debt}
Due to increased review burden, developers reported fatigue and effects on technical debt, but also challenges related to \textit{cognitive} or \textit{intent} debt~\cite{storey2026technical} (undeveloped team codebase understanding or rationale) because of uncertain intent, unobserved merges, and an erosion of nontechnical review benefits. Table~\ref{tab:goal-consequences} shows relevant survey results for this section. 

\begin{table}[h]
\centering
\caption{\textbf{Changes to review outcomes and goals after AI-assisted authoring.} Percentages are respondents who agreed (purple) vs. disagreed (yellow). $n=322$. Survey items are cut for space.}
\label{tab:goal-consequences}
\begingroup
\fontsize{6.6}{7.6}\selectfont
\setlength{\tabcolsep}{0pt}
\setlength{\aboverulesep}{0.1ex}
\setlength{\belowrulesep}{0.1ex}
\setlength{\abovetopsep}{0pt}
\setlength{\belowbottomsep}{0pt}
\renewcommand{\arraystretch}{1.10}
\newcommand{\KaiaBipolarBar}[7]{%
  \begingroup%
  \def\KaiaColorOne{KaiaCompFive}%
  \def\KaiaColorTwo{KaiaCompFour}%
  \def\KaiaColorThree{KaiaCompThree}%
  \def\KaiaColorFour{KaiaCompTwo}%
  \def\KaiaColorFive{KaiaCompOne}%
  \KaiaDistributionBar{#1}{#2}{#3}{#4}{#5}{#6}{#7}%
  \endgroup%
}
\sbox{\KaiaTableBox}{%
\begin{tabular}{@{}>{\raggedright\arraybackslash}p{0.05\columnwidth}@{\hspace{4pt}}>{\raggedright\arraybackslash}p{0.53\columnwidth}@{\hspace{4pt}}>{\raggedright\arraybackslash}p{0.29\columnwidth}@{\hspace{4pt}}>{\raggedleft\arraybackslash}p{\dimexpr\columnwidth-0.05\columnwidth-0.53\columnwidth-0.29\columnwidth-12pt\relax}@{}}
\toprule
\textbf{ID} & \textbf{Change} & \multicolumn{1}{@{\hspace{4pt}}p{0.29\columnwidth}@{}}{\KaiaDistributionHeader{Distribution}} & \textbf{$p_{adj}$} \\
\midrule
DC1 & My review interactions with authors feel less like mentoring opportunities. & \KaiaBipolarBar{0.0931677019}{0.0962732919}{0.1863354037}{0.3602484472}{0.2639751553}{19}{62} & $p < .001$ \\
DC2 & I feel more mentally drained by code review than before. & \KaiaBipolarBar{0.1118012422}{0.1273291925}{0.1987577640}{0.3167701863}{0.2453416149}{24}{56} & $p < .001$ \\
DC3 & I have less patience and focus for code review than before. & \KaiaBipolarBar{0.0993788820}{0.0931677019}{0.2795031056}{0.3354037267}{0.1925465839}{19}{53} & $p < .001$ \\
DC4 & Code review is less valuable for maintaining understanding of the codebase. & \KaiaBipolarBar{0.1863354037}{0.1894409938}{0.1956521739}{0.2795031056}{0.1490683230}{38}{43} & $p = .91$ \\
\bottomrule
\end{tabular}%
}
\usebox{\KaiaTableBox}
\par\vspace{0pt}
\begin{minipage}{\wd\KaiaTableBox}
\fontsize{6.2}{7.2}\selectfont
\KaiaLegend{KaiaCompFive}{Strongly disagree}\quad\KaiaLegend{KaiaCompFour}{Somewhat disagree}\quad\KaiaLegend{KaiaCompThree}{Neither}\quad\KaiaLegend{KaiaCompTwo}{Somewhat agree}\quad\KaiaLegend{KaiaCompOne}{Strongly agree}\\[-4pt]
\noindent\rule{\linewidth}{\heavyrulewidth}
\end{minipage}
\endgroup
\end{table}

\subheading{Fatigue.} Developers reported increased
review burden \textit{impacted their other job responsibilities} (\textbf{C9}, $28\%$, $32\%/21\%$, \textit{weakly more}): \qualquote{So many PRs need to be reviewed. `When do I do my own
work?' is a question that I keep getting from the developers} (P20). Some also reported impacts on their \emph{fatigue} (\textbf{DC2}, $56\%/24\%$, \textit{agree})
or \emph{patience for review} (\textbf{DC3}, $53\%/19\%$, \textit{agree}), e.g, \qualquote{The biggest challenge has been wanting, being asked, to stay totally in
control\ldots 
% but also being
% really disinclined to spend a lot of time on code review\ldots there was a point
% where 
a switch flipped and I no
longer had patience to sit and read line by line} (P19).

\subheading{Debt.} 
AI authoring may influence technical debt: some interviewees perceived an increase in \qualquote{little regressions} (P3) or \qualquote{constant patches}
(P11). AI can also proliferate technical debt:\qualquote{LLMs
view that code [and are] like, `oh, there's an existing pattern here. Let me do that
pattern again'} (P6).
% There are tons of PRs that get through that I don't look
% at \ldots now  

Rapid churn can also lead to \emph{difficulty maintaining codebase understanding}: \qualquote{There are many more PRs\ldots And then
this learning phase, where you actually read the code and understand,
people don't have time for it} (P13). However, this is still a desired benefit of review for many, as surveyed developers did not significantly report a change in how much code review was an opportunity to maintain understanding of the codebase (\textbf{DC4}, $43\%/38\%$, \textit{inconclusive}). Interviewees were reluctant to fully automate review due to a loss of this benefit, e.g., \qualquote{[Code review is] my main mechanism for understanding what's happening in the
product. So that's one reason I'm a little reticent to just turn [my review agent] into an
agent that runs every day and does my work for me} (P21).
This lack of system understanding from churn can lead to more difficulty understanding merge implications: %RQ4
\qualquote{Because of the
rate of change\ldots it's very difficult to know the
interdependencies and\ldots how we make
sure that this is not going to cause any further downstream [impacts]\ldots }
(P20). Rapid churn can even lead to multiple codependent PRs existing in parallel, making merge implications impossible to know (a challenge for P24's team).
Developers reported that code
review is \textit{less of a mentorship opportunity} than before 
(\textbf{DC1}, $62\%/19\%$, \emph{agree}), partly because review interactions sometimes happen via agents: \qualquote{I'm not mentoring the engineers and
building great, best-in-class engineers by
doing code reviews and sharing knowledge
now. I'm just a pass through back to these
LLMs}
% It used to be I'd work with these junior
% engineers\ldots You build these best practices, they
% remember them. And I think we're losing that
(P2). 

% Interviewees reported that although AI was exceptional at catching certain errors, the human reviewer still has an important role: \qualquote{Even if AI gets super good, we still have to approve stuff. A lot of times, the mechanical stuff makes a lot of sense, but does it fit overall with our system?} (P11). 

\summarytcolorbox{Factors from AI authoring, including PR quality issues, productivity pressures, lack of author stewardship, uncertain intent, and PR volume and size, lead to \textbf{increased burden on reviewers.} This burden leads to fatigue and shallow and missed review, which leads to \textbf{downstream debt.}}

\subsection{RQ3 - AI Review Tooling: Impacts and Effective Use}
\label{ai4r}
We detail the \textbf{effects of AI review tooling on review challenges}, \textbf{use cases}, and \textbf{challenges} and \textbf{strategies} for effective use. Participants often used general-purpose AI or ``skills'' (i.e., reusable prompts) and platforms that they or a team maintained, in manual or autonomous ways. 

\subsubsection{Effects on Review Volume}
\label{airevieweffects}
Difficulty keeping up with the volume and pace of review made using AI for review \textit{necessary} for many participants: \qualquote{That's actually why I wrote the agent\ldots all of a sudden it was like, oh, wow, I could be spending 8 hours a day on nothing but code reviews because there's so many coming in so fast and they're so big} (P21).
% \jenna{not sure you need the work whether here} 
% kln: i believe we do: we tested whether three measures of adoption....were correlated with X. the nested whether does look weird though
To investigate whether using AI for review mitigates review volume difficulty, we tested whether four measures of AI use and exposure---whether respondents had used AI for review or seen AI-generated comments on PRs they review, and how frequently they use AI for review or encounter AI-generated comments (among those who had done so in the past six weeks)---were correlated with current review volume difficulty or changes in that difficulty after AI-assisted authoring.
No adoption measure was correlated with a \textit{decrease} in current difficulty or with a perceived change in difficulty after AI-assisted authoring
(Spearman's $\rho$, Mann-Whitney $U$). Personally using AI for review at all was instead correlated with \textit{increased} current difficulty ($p = .006$, $r=.16$) with a small effect size, as was more frequently seeing AI-generated comments (a trend at $p = .09$, $\rho=.08$). Although these correlations cannot establish causation, interview evidence suggests that increased PR volume may motivate adoption of AI for review, rather than adoption causing difficulty with volume. These results \textbf{do not support a narrative that using AI for review eliminates AI-driven review volume difficulties.} 
These results do not reflect low adoption: of all respondents ($n=2{,}590$), $93\%$ have seen an AI review bot
comment on PRs they review, $69\%$ have seen other AI-generated comments, and
$82\%$ have personally used AI for review, $80\%$ of whom ($n=2{,}093$) use it
at least half the time. Only $1\%$ had no AI review experience.
% $82\%$ of all respondents ($n=2{,}590$) have personally used AI for review, and $80\%$ of adopters ($n=2{,}093$) reported using it at least half the time.
However, survey respondents thought \textbf{AI had improved the quality and speed of their review} ($78\%$ better vs. $4\%$ worse quality; $68\%$ faster vs. $11\%$ slower). Some interviewees' experiences with AI review were unhelpful (e.g., P2) or mixed (e.g., P6).
% e.g., \qualquote{It's not saving me a ton of time\ldots it's now increased my quality a little bit} (P6).

\subsubsection{AI Review Use Cases}
\label{aireviewusecases}
\begin{table}[t]
\caption{\textbf{Current AI use for review} ($n=996$) and \textbf{responses to AI-generated
comments} ($n=1202$), by \% of respondents. Survey items shortened for space.}
\label{tab:ai-usage}
\fontsize{6.8}{7.8}\selectfont
\renewcommand{\arraystretch}{0.95}
\setlength{\tabcolsep}{3pt}
\begin{minipage}[t]{0.53\linewidth}
\begin{tabularx}{\linewidth}{lXr}
\toprule
\textbf{ID} & \textbf{Current AI use} & \textbf{Used\%} \\
\midrule
\rowcolor{rowColor} U1 & Understand the PR better and guide review & 79\% \\
U2 & Discuss to refine and understand comments & 67\% \\
\rowcolor{rowColor} U3 & Generate comments and selectively post & 46\% \\
U4 & Generate comments and edit before posting & 39\% \\
\rowcolor{rowColor} U5 & Check that previous comments were addressed & 34\% \\
U6 & Prioritize changes within a PR & 25\% \\
\rowcolor{rowColor} U7 & Generate comments and auto-post them as-is & 23\% \\
U8 & Prioritize generated comments & 15\% \\
\rowcolor{rowColor} U9 & Orchestrate other review agents with an agent & 14\% \\
U10 & Prioritize PRs & 13\% \\
\rowcolor{rowColor} U11 & Generate comments and manually post as-is & 13\% \\
U12 & Other & 7\% \\
\bottomrule
\end{tabularx}
\end{minipage}\hfill
\begin{minipage}[t]{0.44\linewidth}
\begin{tabularx}{\linewidth}{lXr}
\toprule
\textbf{ID} & \textbf{Response to AI comments} & \textbf{Used\%} \\
\midrule
\rowcolor{rowColor} CI1 & Checked whether valid or relevant & 89\% \\
CI2 & Used to understand PR or guide attention & 57\% \\
\rowcolor{rowColor} CI3 & Dismissed or flagged unhelpful ones & 56\% \\
CI4 & Endorsed or elaborated on useful ones & 51\% \\
\rowcolor{rowColor} CI5 & Followed up with authors & 47\% \\
CI6 & Other & 4\% \\
\rowcolor{rowColor} CI7 & Mostly ignored them & 4\% \\
\bottomrule
\end{tabularx}
\end{minipage}
\end{table}
Table~\ref{tab:ai-usage} shows data on how survey respondents used AI for review and their interactions with other AI-generated comments on PRs they review.

\subheading{Understand PRs.} Developers used AI to \textit{understand PRs} or guide their review (\textbf{U1}, used by $79\%$). Interviewees used AI to ask \qualquote{pointed questions} (P18) about common or potential issues, get PR summaries or line-by-line walkthroughs, or interrogate and contrast information sources. Interviewees (e.g., P22) used AI to understand PRs more often when in an \textit{unfamiliar} part of the codebase or when PRs were \textit{large or complex}.
% Interviewees (e.g., P9) also noted AI had decreased the burden on the author to explain their PR. 
% above is interesting but isn't justified by space
Developers also check whether AI comments posted on PRs they review are valid (\textbf{CI1}, $89\%$) or use them to understand PRs (\textbf{CI2}, $57\%$).

\subheading{Generate Comments.} Developers used AI to \textit{generate review comments} for PRs (used by $39-46\%$ respondents (\textbf{U3}, \textbf{U4}) in a human-in-the-loop process, and $13-23\%$ (\textbf{U7}, \textbf{U11}) autonomously). Developers edit, augment, omit, verify, and \textit{discuss} (\textbf{U2}, $67\%$) review comments to \textit{improve comment precision or presentation} and \textit{improve issue understanding}. For example, P19 would sometimes use the code location ``anchored'' to an AI-generated comment to verify it: \qualquote{I can either have the pull request open\ldots and look at that anchor position and come to my own conclusion, or I can ask follow-up questions\ldots it [asks] me to approve, reject, skip, or discuss.}

\subheading{Prioritize Attention.} Developers used AI to \textit{prioritize their attention between changes} (\textbf{U6}, $25\%$), \textit{between comments} (\textbf{U8}, $15\%$), and \textit{between PRs} (\textbf{U10}, $13\%$). Interviewees reported that AI prioritization often relied on issue priority or severity, supporting the attentional strategies described in Section~\ref{reviewerattention}. Interviewees (e.g., P18) also used AI to show how changes fit together (e.g., based on their control flow or dependencies) to guide their attention and understanding. 

\subheading{Orchestrate Review Agents.} Developers used \textit{agent orchestration} (\textbf{U9}, $14\%$) to have AI agents delegate review agents to certain PRs or multiple to one (in the case of multiagent systems). For example, P21 used agent orchestration to aggregate and rank generated comments across PRs into a report, which he mediates individually. P19 had orchestration agents dynamically dispatch resources used (e.g., LLM models, tokens, numbers of agents, etc.) based on review priority. 
 
\subheading{Track Iterations.} Developers used AI to check whether \textit{comments had been integrated} (\textbf{U5}, $34\%$) after iteration, e.g., \qualquote{If\ldots the developer comes back and says, `oh, I've fixed all the comments'\ldots [AI] looks at all the responses\ldots and check[s]\ldots if the fix is actually accurate} (P20). Developers also tracked iterations by checking author responses to AI-generated comments (\textbf{CI5}, $47\%$). 
% \jenna{Maybe a summary line here about again, with all of these strategies, it still isn't related to reduced challenges?}
% kln: little tight on space, but i will add this in post 

\subsubsection{AI Review Tooling Challenges and Effective Strategies} 
\label{aireview challenges}
We detail \textbf{six categories of AI review tooling challenges}, with indicative challenges and strategies shown in Table~\ref{tab:ai-review-chal-strat}. 
% Developers needed to customize and supervise AI review tools: only 
% $2\%$ reported they worked off-the-shelf.

\subheading{Precision.} The most frequent surveyed AI review challenge was \textit{incompleteness} of AI comments (\textbf{AC1}, $54\%$ reported occurring half the time or more). Survey results also indicate \textit{nitpicking comments} (\textbf{AC5}, $42\%$), \textit{comment inaccuracy} (\textbf{AC7}, 35\%), \textit{repetitive mistakes} (\textbf{AC10}, $28\%$), and \textit{repetitive mistakes despite prior instruction} (\textbf{AC15}, $17\%$) . Unsound comments may also be \textit{subtly incorrect}: \qualquote{[AI] makes a comment that actually makes no sense\ldots  but superficially it might make sense and people go ahead and change the code to do it that way and introduce problems} (P10).
Developers reported that incorrect AI review feedback can result from a lack of \textit{implicit project knowledge} (\textbf{AC8}, $32\%$) or \textit{code or documentation beyond the PR} (\textbf{AC9}, $28\%$).

\begin{table*}[t]
\centering
\captionsetup{font=scriptsize, labelfont=scriptsize}
\caption{\textbf{Selected AI review challenges and corresponding strategies.} \textit{Top}: challenges. \textit{Bottom}: indicative strategies for each challenge category (some had no strategy), ranked by a Plackett-Luce model fit to respondents' top two most helpful
strategies among those they had used or seen used. Worth is the estimated probability that a
strategy is ranked first.
ST = social tension, UE = usage effort, OR = overreliance, PM = perception management, PR = precision, PS = presentation.}
\label{tab:ai-review-chal-strat}
\begingroup
\fontsize{7.2}{8.0}\selectfont
\setlength{\tabcolsep}{0pt}
\setlength{\aboverulesep}{0.1ex}
\setlength{\belowrulesep}{0.1ex}
\setlength{\abovetopsep}{0pt}
\setlength{\belowbottomsep}{0pt}
\renewcommand{\arraystretch}{1.0}
\sbox{\KaiaTableBox}{%
\begin{tabular}{@{}>{\raggedright\arraybackslash}p{0.06\textwidth}@{\hspace{6pt}}>{\raggedright\arraybackslash}p{0.52\textwidth}@{\hspace{6pt}}>{\raggedright\arraybackslash}p{0.02\textwidth}@{\hspace{6pt}}>{\raggedright\arraybackslash}p{\dimexpr\textwidth-0.06\textwidth-0.52\textwidth-0.02\textwidth-18pt\relax}@{}}
\toprule
\textbf{ID} & \textbf{AI Review Challenge} & \textbf{Cat.} & \multicolumn{1}{@{\hspace{6pt}}p{\dimexpr\textwidth-0.06\textwidth-0.52\textwidth-0.02\textwidth-18pt\relax}@{}}{\KaiaDistributionHeader{Distribution}} \\
\midrule
AC1 & Not all important issues were caught by AI comments & PR & \KaiaFrequencyBar{0.0487603306}{0.4123966942}{0.2520661157}{0.2322314050}{0.0545454545}{46}{54} \\
AC2 & AI comments were overly verbose or complex & PS & \KaiaFrequencyBar{0.1267605634}{0.3405136703}{0.1748135874}{0.2336371168}{0.1242750621}{47}{53} \\
AC3 & I worried my own review was insufficient after relying on AI & OR & \KaiaFrequencyBar{0.2242798354}{0.3127572016}{0.1759259259}{0.1820987654}{0.1049382716}{54}{46} \\
AC4 & AI-generated comments were not engaged with by the author & PM & \KaiaFrequencyBar{0.1489184692}{0.3968386023}{0.2212978369}{0.2121464226}{0.0207986689}{55}{45} \\
AC11 & AI review feedback was difficult or time-consuming to verify & UE & \KaiaFrequencyBar{0.2654320988}{0.4578189300}{0.1676954733}{0.0895061728}{0.0195473251}{72}{28} \\
AC14 & PR authors expressed frustration with AI comments & ST & \KaiaFrequencyBar{0.4104289319}{0.4003364172}{0.1135407906}{0.0630782170}{0.0126156434}{81}{19} \\
\bottomrule
\end{tabular}%
}
\usebox{\KaiaTableBox}
\par\vspace{0pt}
\begin{minipage}{\wd\KaiaTableBox}
\fontsize{6.2}{7.0}\selectfont
\KaiaLegend{KaiaFreqOne}{Never}\quad\KaiaLegend{KaiaFreqTwo}{Sometimes}\quad\KaiaLegend{KaiaFreqThree}{About half the time}\quad\KaiaLegend{KaiaFreqFour}{Most of the time}\quad\KaiaLegend{KaiaFreqFive}{Always}\\[-4pt]
\noindent\rule{\linewidth}{\heavyrulewidth}\\[1pt]
Per-item $n=963$--$1214$. Numbers on the right of the distribution indicate \% ``About half the time'' or more; those to the left indicate \% ``Sometimes'' or ``Never''.
\end{minipage}
\endgroup

\vspace{4pt}

\scriptsize
\setlength{\tabcolsep}{2pt}
\setlength{\aboverulesep}{0.1ex}
\setlength{\belowrulesep}{0.1ex}
\setlength{\abovetopsep}{0pt}
\setlength{\belowbottomsep}{0pt}
\renewcommand{\arraystretch}{1.0}
\begin{tabularx}{\textwidth}{@{}l X l l l l@{}}
\toprule
\textbf{ID} & \textbf{AI Review Strategy} & \textbf{Cat.} & \textbf{Used \%} & \textbf{Worth} & \textbf{CI} \\
\midrule
\rowcolor{rowColor} AIS2 & Have AI review do multiple passes or put multiple agents on one PR & PR & 55\% & 0.20 & [0.14, 0.25] \\
AIS3 & Add repository- or team-specific review expectations to reusable instructions (e.g., a skill) & PR & 55\% & 0.11 & [0.08, 0.15] \\
\rowcolor{rowColor} AIS10 & Encourage developers to engage with AI review comments & PM & 48\% & 0.02 & [0.01, 0.02] \\
AIS11 & Scale back AI review based on negative developer feedback & ST & 11\% & 0.02 & [0.00, 0.03] \\
\rowcolor{rowColor} AIS13 & Instruct AI review to adjust the length, tone, or format of comments & PS & 26\% & 0.01 & [0.00, 0.01] \\
\bottomrule
\end{tabularx}

\vspace{1pt}
{\raggedright\scriptsize
$n=1203$ respondents indicated whether they used each strategy; 3\% selected Other, 3\% selected ``None, despite challenges,'' and 2\% selected ``None, works off-the-shelf.'' $n=942$ respondents (those whose top favorites were not Other) were used to fit the Plackett-Luce model.\par}
\end{table*}

To improve AI review precision, developers reported using skills with \textit{repository-} or \textit{team-specific instructions} (\textbf{AIS3}, used by $55\%$)
and would \textit{customize instructions} \aistrat{9}{39} for different repositories and areas of the code. Some interviewees (e.g., P7) used reusable prompts maintained by developer tooling teams and changed instructions for their use cases. These instructions were noted to benefit both AI review and AI authoring, e.g., \qualquote{We have instructions that are specific to an app\ldots we also want the people to use [the] same instructions when they build the code} (P13). Developers also gave AI review \textit{access to code or documentation} \aistrat{4}{46}. AI review precision was improved in iterative processes where developers would \textit{incorporate human feedback} from review sessions into AI review instructions or context \aistrat{12}{25}, e.g., \qualquote{I don't have 100\% trust that [AI] has got everything\ldots We've always understood that \textbf{the human in the loop is the thing that is helping us make it better over time.} The efficacy of review is improving because the human in the loop is actually tweaking the instructions} (P20).
% \qualquote{If you are finding that you're
% leaving the same comment on all the code reviews, we need to go change the AI
% instructions\ldots Otherwise, we're just playing whack-a-mole here}
% (P1). 

Developers also \textit{adjusted prompt tone} \aistrat{8}{34} to improve precision. \textit{Multi-agent} or -pass systems were also perceived as highly useful for AI review precision \aistrat{2}{55}, though interviewees (e.g., P22) noted that multiple agents or passes would sometimes catch the same issues. To raise the potential for diversified AI review findings and or catching critical issues, developers also \textit{specified concerns} for AI review \aistrat{6}{47}, e.g., \qualquote{I have very specific focus area[s]\ldots security is important, or performance is important\ldots I know each individual person's behavior over the last some years of time with them\ldots so [I ask] AI to pull that out} (P22). Developers also mediated AI-generated comments during review to improve precision and reduce social tension.

\subheading{Presentation.} Developers reported \textit{overly verbose or complex comments} (\textbf{AC2}, $53\%$), \textit{overly-numerous comments} (\textbf{AC6}, $39\%$), \textit{duplicate comments} (\textbf{AC16}, $16\%$), and \textit{ill-formatted comments} (\textbf{AC17}, $12\%$).

Developers adjusted how AI review findings were displayed to reduce social tension with authors and the review burden on themselves. Developers instructed AI review to \textit{adjust tone and reduce verbosity} of comments \aistrat{13}{26} and \textit{configured the display of review findings} in custom applications or command line output for themselves \aistrat{14}{10} to reduce cognitive load, e.g., \qualquote{It will present [comments] to me one at a time\ldots I cannot tolerate having to scroll back through 15 different things and read them all at the same time and make my choices} (P19). 

\subheading{Social Tension.} Developers reported that authors could get \textit{frustrated} with poor AI-generated comments (\textbf{AC14}, $19\%$), e.g., \qualquote{an agent would go rogue\ldots and shotgun blast somebody's pull request with a bunch of comments [that] may or may not be legitimate\ldots I was embarrassed when it happened on my
behalf, and I have a tendency to take it personally when it happens to me} (P19).

If automated AI review was not precise enough or poorly presented, developers sometimes \textit{scaled back AI review} \aistrat{11}{11} to reduce ire from developers involved in the review process, e.g., \qualquote{It was autonomous for a little bit, and it went nuts\ldots I got feedback from people like, `hey, this thing is way too verbose.'\ldots 
% I tried reducing the verbosity and it was still too long, so 
at this point it is a bot that I run locally\ldots [and] sift through the feedback\ldots} (P6).

\subheading{Perception Management.} Interviewees had trouble getting authors to attend to, trust, and properly evaluate AI-generated comments, e.g., \qualquote{When we run these agentic reviews, 
% especially for partner teams, 
it's important for them to recognize that this is a first pass\ldots people will see agentic reviews and they'll be like, `oh, okay, no human reviewed it'} (P1). Previous experiences with AI review can also influence perception: \qualquote{The first iterations of the code review bot were super noisy and so people are going to ignore it} (P2).
Surveyed developers frequently saw AI-generated comments \textit{without author interaction} (\textbf{AC4}, $45\%$), and sometimes saw unanswered comments when
they felt a response was \textit{necessary} (\textbf{AC13}, 23\%).
% \emerson{Could we clarify ``no interaction'' versus ``without needed interaction''? I take the first to mean nobody responded to the comment and the second to mean nobody responded even though the respondent thought a response was warranted. If that is what the survey items mean, please say so explicitly.}
% Interviewees (e.g., P10) also noted poor AI review comments, if posted under the reviewer's name, could cause authors to distrust their comments in the future. 
% I think the biggest one is it like there's a lack of trust there that if I'm putting my name on it, that means that I trust it to comment on stuff the way I would do it. And if it doesn't actually do that, then people like people in my team might start doubting like legitimate comments that I live in the PRs.
% So that's one concern.
% for the above on perception management, inlining this quote could be more expedient than writing all that shit out
% this is interesting, but not sure if it deserves the space

To manage author perception of AI comments, developers reported dismissing or \textit{flagging} inaccurate comments (\textbf{CI3}, $56\%$, see Table~\ref{tab:ai-usage}), \textit{endorsing} accurate ones (\textbf{CI4}, $51\%$), and \textit{following up} with authors about comments (\textbf{CI5}, $47\%$). Developers also reported \textit{encouraging developers to engage} with AI-generated review comments \aistrat{10}{48}. Interviewees (e.g., P18) also reported strategically avoiding letting review agents post autonomously or leaving a note that they personally reviewed the code to assure authors there was a human reviewer involved.

\subheading{Usage Effort.} AI review can be \textit{difficult to verify} (\textbf{AC11}, $28\%$) and \textit{effortful to configure} (\textbf{AC12}, $26\%$). Interestingly, some interviewees did not personally use AI for review because it goes beyond their responsibility: \qualquote{Invoking review on someone else's PR\ldots \textbf{I am taking on the responsibility of verifying all of the things that the AI says} also, versus if the [bot] comments\ldots then that is the reviewer's responsibility to figure out\ldots} (P12). 
% there's a lot that I might have to consider that would have been way out of scope} (P12). 

\subheading{Overreliance.} Surveyed developers were often worried about \textit{overreliance} on AI leading to shallower review (\textbf{AC3}, $46\%$). Interviewees were actively determining what level of engagement is necessary, especially for nontechnical review value: \qualquote{A majority of code changes become more black box, because I am asking AI to translate it\ldots I'm trying to figure out what's the best way I still can\ldots understand the overall picture and also details of the changes} (P22). 
% Reliance on AI could also impact : 
% \qualquote{If my agents are looking at all the code for me, I don't get any of those
% secondary benefits, right? How can I keep myself in the loop and engaged in
% a way that I still get those secondary benefits?} (P19). 
Some interviewees (e.g., P11) avoided use of agentic review due to concerns about cognitive debt.
% P24: We actually think our code review process is a bit broken in the team right
% now. So that's, we think that we are shipping way too many Prs and not enough
% people are even looking at. People are just 100\% relying on a code review from
% AI now, as opposed to, there's a part to play for the humans still, right?

\summarytcolorbox{AI-driven PR \textbf{volume motivates reviewers to use AI for review}, but adoption \textbf{does not correlate with decreased difficulty} keeping up with review volume. Developers primarily used AI to understand PRs, generate comments, and prioritize attention during review. Developers reported the most severe challenges when using AI for review are precision, presentation, overreliance, and perception management, and offered helpful strategies for all but overreliance.}
\subsection{RQ4 - Strategies Mitigating Review Challenges}
\label{strats}
Developers reported strategies based in \textbf{organizational practices}, \textbf{review workflows}, and \textbf{reviewer attention} that are helpful to resolve increased review burden from AI authoring and \textbf{facilitate an effective shift toward high-level review} for human reviewers.
Surveyed strategies developers used are ranked by perceived helpfulness in Table~\ref{tab:gen-strats}.

\begin{table}[h]
\centering
\captionsetup{font=scriptsize, labelfont=scriptsize}
\caption{\textbf{Non-AI strategies for code review, by perceived helpfulness.} Helpful \% is the \% who picked each strategy among their top two most helpful. Used \% is the \% who have used strategies or seen them used.}
\label{tab:gen-strats}
\scriptsize
\setlength{\tabcolsep}{2pt}
\renewcommand{\arraystretch}{1.0}
\begin{tabularx}{\columnwidth}{@{}l X l l@{}}
\toprule
\textbf{ID} & \textbf{Strategy} & \textbf{Used \%} & \textbf{Helpful \%} \\
\midrule
\rowcolor{rowColor} S1 & Create expectation for self-review step before submitting a PR (e.g., with a review skill or manually) & 41\% & 27\% \\
S2 & Give reviewers more context on the author's intent for the PR or decisions made in it & 36\% & 26\% \\
\rowcolor{rowColor} S3 & Improve documentation of implicit knowledge about the codebase & 40\% & 19\% \\
S4 & Have organization or management incentivize or encourage code review & 27\% & 16\% \\
\rowcolor{rowColor} S5 & Require PRs to be ready before requesting review (e.g., passing build, won't be iterated on, etc.) & 41\% & 15\% \\
S6 & Improve or create a standardized pipeline for routing PRs to reviewers & 26\% & 12\% \\
\rowcolor{rowColor} S7 & Block PRs over a certain size or complexity (e.g., touching too many files) & 18\% & 11\% \\
S8 & Invest more resources (e.g., money, formal teams) in code review tooling or processes & 19\% & 11\% \\
\rowcolor{rowColor} S9 & Assign more people or hours for code review & 31\% & 11\% \\
S10 & Give reviewers more context on how the PR was made (e.g., if/how AI was used) and the author's understanding & 18\% & 10\% \\
\rowcolor{rowColor} S11 & Have organization or management measure and prioritize code quality & 24\% & 10\% \\
S12 & Provide code review training or dedicated sessions for sharing knowledge and tools & 12\% & 10\% \\
\rowcolor{rowColor} S13 & Improve or create tools or processes for prioritizing which PRs need review first & 17\% & 8\% \\
S14 & Meetings or messaging about community norms to improve code review culture & 22\% & 3\% \\
\bottomrule
\end{tabularx}

\vspace{2pt}
\scriptsize
Used \% $n=646$; Helpful \% $n=635$. $5\%$ used Other ($3\%$ favored), $9\%$ selected ``None, we haven't done anything to help with code review challenges,'' and $4\%$ selected ``None, we have no significant code review challenges''.
\end{table}

\subsubsection{Organizational Practices}
\label{orgpractices}
To mitigate increased review burden from PR quality issues and productivity pressures, developers thought the most helpful strategies involved organizations \textbf{promoting review and improving its conditions.}

\subheading{Document Implicit Knowledge.} To align developers and AI tooling and better scale their
capabilities, developers found \emph{documentation of implicit knowledge}
about the codebase helpful (\textbf{AIS5}, $48\%$; \textbf{S3}, used or seen used by $40\%$). \qualquote{That's [what] I'm most looking forward to, to have
well-defined what used to be tribal knowledge\ldots 
so the agents can not only review that, but also [it] 
can be available when the agent is producing the code to start
with} (P9). Documentation and
curation can be difficult, as AI can pick up on unintentional patterns, e.g., 
\qualquote{Sometimes [AI agents] are trying to adhere to these decisions that
were never intended to be that strict. And then [documentation] ends up creating more
friction rather than less} (P19). 
% Interviewees (e.g., P23) used retrieval-augmented generation to aid this.
% Additionally, if not prioritized or supported, interviewees (e.g., P16) noted that other engineering priorities can take precedence over
% writing documentation.
% \qualquote{Our team has put together our own robot\ldots I have a to-do
% item\ldots about writing about how our system fits
% together and how data flows\ldots I
% don't have budget to do it right now. I've talked with my manager and 
% I need to finish a couple assignments\ldots} (P16).

\subheading{Resource and Promote Review.} 
% sadly i think it's a waste of space to include the ``thread of sad people
% crying'' part of this quote but it amused me greatly
Developers thought having organizations prioritize and \emph{incentivize
code review} \genstrat{4}{27}, \textit{devote resources to code review} \genstrat{8}{19}, or \textit{assign more reviewers or hours} \genstrat{9}{31} would be particularly helpful for decreasing
review latency, e.g., \qualquote{If there was an
actual organizational push for people to be doing reviews, [review] would be going
faster. Apparently everyone wants things to go super fast here, so I don't
understand why that's still a problem} (P15). It may also aid author stewardship and reduce cognitive debt: \qualquote{The issues
that we've been seeing with people putting up all these changes that they
don't really understand\ldots wouldn't be as large of
an issue if people were regularly doing code review} (P12). 
Developers also reported attending
\emph{meetings to improve code review culture} \genstrat{14}{22} or dedicated \emph{knowledge and tool sharing sessions} \genstrat{12}{12}. 

\subheading{Measure Review Quality.} Developers thought organizations could \textit{measure code quality} \genstrat{11}{24} or review health to better assess engineering productivity and mitigate productivity
pressures, e.g., \qualquote{Speed is what's being measured\ldots it
doesn't seem like we're measuring how many iterations on these PRs are we going
through\ldots 
% How much time from senior engineers are we spending?\ldots 
If we [measure] the right things\ldots things would get better} (P6). 

\subsubsection{Review Workflows}
\label{prpreparation}
To mitigate increased review burden from PR quality issues, uncertain intent, and a lack of author stewardship, developers thought the most helpful strategies \textbf{push review burden upstream to authors and tooling.}

\subheading{Pre-Review.} Requiring a step where \textbf{authors or AI tools
review PRs before human review} was \textbf{the most helpful strategy} across our survey questions on organization and workflow as well as AI review strategies (\textbf{S1}, $41\%$; \textbf{AIS1}, $69\%$). Interviewees noted that pre-review could clear up AI-driven PR quality issues 
 or consequences of diminished author
stewardship:
\qualquote{I feel like it should be mandatory\ldots you must carefully review your
own code before you're asking my time to review the code the robot wrote} (P16). 
Interviewees also described their
pipelines which had \textit{automated kick-off} \aistrat{7}{38} or 
expectations for running centrally- or personally-maintained AI review skills
prior to submission, e.g., 
% For example, P22, an engineering manager, described how he
% checks to make sure engineers run a centrally-maintained AI review tool: 
\qualquote{We wrote our own code review agent\ldots I ask everyone to run it before they
create the PR\ldots I repeat the same question through
AI on the [DevOps platform] and see whether they actually did the code review
before or not} (P22, an engineering manager). 
However, publicly posted AI-generated comments may clutter PRs for human reviewers: \qualquote{The agents really love to
comment on the entire file, which makes it impossible to review\ldots} (P8). 

\subheading{Require Merge Readiness.} Developers thought \emph{requiring PR readiness} before requesting human review \genstrat{5}{41} could reduce review burden. Developers (e.g., P23) also found it helpful to \emph{block} \genstrat{7}{18} or require authors to chunk PRs with large
changesets, a common best practice~\cite{macleod2017code}. This gating of PRs,
along with pre-review steps by authors and AI review, were seen as strategies 
that could \textbf{free reviewers up to focus on the
high-level concerns that now define the human reviewer's role} (see
Section~\ref{quovadis}):
 \qualquote{I think we're missing a step\ldots 
 if we spend time reviewing something that\ldots 
will be iterated on 10 times based on agent review comments\ldots It's a problem to be solved to \textbf{have a
clear way to signal that\ldots this is
ready to review for humans}\ldots All the common mistakes that you have encoded as review agents are out of the
way, so \textbf{now you can focus
100\% on the big picture}} (P9).
% and the thing that is hard to encode in a markdown}
Interviewees (e.g., P3) also marked PRs as drafts
to signal they were not stable.

\subheading{Document Intent and Provenance.} Giving reviewers \emph{context on author intent} for the PR
\genstrat{2}{36} could help reviewers evaluate whether PRs
follow their desired purpose and design. For example, P22 asked his
team to \qualquote{create a plan
doc\ldots explain what's the design, why did I do it, what
pattern I'm taking\ldots 
and then keep the plan along with your change, so people can easily
approach your code.} 

\emph{Context on how PRs are made} and author understanding \genstrat{10}{18} could also help reviewers better calibrate review scope and
depth. 
If reviewers know the provenance of a change, they can selectively look
for human-typical errors:
\qualquote{If I know that I'm reviewing
human-authored code, I'm going to check every conditional
and make sure that you got the logic right} (P14). 
They can also selectively look for AI-typical errors (e.g., loss of intent):
\qualquote{Leaders in the org wanted people to disclose when they had 
 used AI\ldots [and] if AI did the whole thing\ldots It kind of totally changes how you look at the PR in terms of, 
 `Am I also trying to figure out if this change is the right change at all?'} (P12) 
However, some interviewees
noted it is difficult to label the degree to which a PR is ``AI-authored'' (e.g.,
P5) or that in practice, they do not find AI disclosures useful (e.g., P19). 

\begin{table}[h]
\centering
\captionsetup{font=scriptsize, labelfont=scriptsize}
\caption{\textbf{Attention allocation strategies, ranked by perceived helpfulness.} Worth is the estimated probability that a strategy is ranked first.}
\label{tab:attention-strats}
\scriptsize
\setlength{\tabcolsep}{2pt}
\renewcommand{\arraystretch}{1.0}
\begin{tabularx}{\columnwidth}{@{}l X l l l@{}}
\toprule
\textbf{ID} & \textbf{Option} & \textbf{Used \%} & \textbf{worth} & \textbf{CI} \\
\midrule
\rowcolor{rowColor} AS1 & Being selective or prioritizing things I pay attention to within a PR when reviewing & 57\% & 0.16 & [0.12, 0.20] \\
AS2 & Allocating more time for code review & 34\% & 0.16 & [0.11, 0.21] \\
\rowcolor{rowColor} AS3 & Scoping down the domain of PRs I review (e.g., to a smaller part of the codebase) & 30\% & 0.15 & [0.10, 0.20] \\
AS4 & Assessing risk to calibrate how much attention to pay to a PR & 36\% & 0.13 & [0.09, 0.18] \\
\rowcolor{rowColor} AS5 & Triaging my review queue based on importance or where I can add the most value & 40\% & 0.11 & [0.08, 0.14] \\
AS6 & Assessing a PR up front to decide whether to review it in depth, wait on author, or skip it & 41\% & 0.10 & [0.07, 0.13] \\
\rowcolor{rowColor} AS7 & Being selective about comments I leave (i.e., avoiding commenting on less major issues) & 38\% & 0.10 & [0.07, 0.13] \\
AS8 & Doing related code reviews together to decrease unnecessary context switching & 21\% & 0.09 & [0.05, 0.13] \\
\bottomrule
\end{tabularx}

\vspace{2pt}
\scriptsize
$n=616$ respondents indicated whether they used each strategy; 13\% selected Other, and 8\% selected ``I don't do any of these more than I used to before.'' $n=434$ respondents (those whose top favorites were not Other) were used to fit a Plackett-Luce model to rank strategies.
\end{table}
\subsubsection{Reviewer Attention}
\label{reviewerattention}
To mitigate attention strain from PR volume and size, developers reported prioritizing their attention during review (see Table~\ref{tab:attention-strats}), often using AI to help. Developers reported that pressure from PR volume \textit{necessitates} a shift toward reviewing for higher-level concerns.

\subheading{Within-PR Attention Strategies.} Developers scope their review to reduce within-PR attention strain burden by \emph{prioritizing concerns} within PRs \attentionstrat{1}{57} 
and being \emph{selective about comments
made} \attentionstrat{7}{38}, e.g., \qualquote{[I review] more PRs on the top level\ldots I need to pick what I'll focus
on} (P13).
Developers
indicated that they \emph{assess PR risk or importance} to determine the depth of
review needed \attentionstrat{4}{36}.
% For example, P9 indicated that he \qualquote{isolate[s] the critical
% parts of the PR and focus[es] firsthand only on this.}
To aid context switching, some developers noted that they
complete related reviews together to reduce onboarding effort \attentionstrat{8}{21}. 

\subheading{Between-PR Attention Strategies.} Reviewers adapted to AI-driven PR volume by \emph{allocating more time for code review} \attentionstrat{2}{34}.
% e.g., \qualquote{It's engineering etiquette: a pull request has
% to contain a single functional change\ldots 
% I want to concentrate on the interesting part} (P23).
To improve review routing, developers (e.g., P2) adapted their attention by \emph{scoping the domain} of PRs
they review \attentionstrat{3}{30}. Developers also
reported \emph{assessing PRs up front} \attentionstrat{6}{41} based on factors such as the depth and
quantity of other reviews or PR quality to
determine if they needed their review yet, if at all:
\qualquote{If folks have build errors, I'll wait\ldots My threshold for reviewing a PR now is higher because the volume has
increased. I want to use my time to review PRs that are ready to be merged} (P3).
Similar to how developers prioritize between changes within a PR, developers
also reported \emph{triaging review queues} \attentionstrat{5}{40} based on factors like risk,
importance, deadline proximity, and the relative value they had to add, e.g.,
\qualquote{I can often tell\ldots whether it's important or not\ldots I know this guy works on these critical areas, so if
he's got a pull request out, then it's probably important to look at it} (P21). To prioritize attention between PRs, developers also thought tooling and workflows could be adapted to \emph{improve or standardize review routing} \genstrat{6}{26} or 
\emph{prioritize between PRs} \genstrat{13}{17}.

\summarytcolorbox{The strategies reported as most helpful for mitigating review burden and potentially freeing up reviewers to do higher-level review involve \textbf{organizations promoting review and improving its conditions} (e.g., by documenting implicit knowledge for AI tooling) and \textbf{shifting review work upstream to authors and tooling} (e.g., through a pre-review step).}

\section{Limitations and Divergent Findings}
\label{limitations}

\subheading{Limitations.} Reported helpfulness of strategies is not demonstrated effectiveness. Future work could establish whether strategies participants found helpful measurably improve review outcomes. Our study also captures an uncertain transition from manual review toward AI-supported workflows; practices reported are provisional, rather than settled accounts of the future.
Few survey respondents ($n=209$, $8\%$) had
only reviewed post-AI, so the experiences of ``AI-native''
reviewers, including whether they build the same codebase knowledge, remain
largely unknown.

\subheading{Threats to Validity.} Survey items may not measure themes expressed by interview participants, threatening \textit{construct validity}~\cite{ralph2018construct}. We mitigate this by providing quotes and survey item wording so readers may judge their agreement. Comparison questions also had lower sample sizes, so some inconclusive results may not signal non-generality of phenomena.

\subheading{Interview and Survey Divergences.} Interview and survey responses sometimes diverged. Divergence may reflect sample sizes or construct invalidity, but it also reveals tensions worth exploring. 
% We discuss challenges that interviewees described as worsened but that surveyed developers reported as significantly less frequent.

On \textit{author stewardship}, interviewees reported more recurring mistakes in AI-authored PRs, less respect from authors, and more revisions ignoring reviewer concerns, while surveyed results significantly show the opposite (\textbf{C15}, \textbf{C18}, \textbf{C21}, \textbf{C20}). Participant attributes may partially explain this gap: interviewees were senior developers selected for high review load, whereas survey respondents were randomly sampled. Developers with high review load may encounter challenges with author stewardship more often, increasing review burden.
One repository we recruited interviewees from also contains many legacy codebases, where AI was reported to be imprecise (see Section~\ref{r4aichal:productivitypressures}), which may shift more burden onto reviewers. Future work could investigate which reviewer attributes and sociotechnical conditions influence AI-driven review challenges.

On \textit{test and description quality}, interviewees also reported that AI-generated PR descriptions often omitted \textit{why} changes were made and that AI-generated tests could be vacuous or overfit (commonly-surfaced issues~\cite{konstantinou2024llms}), whereas surveyed developers reported lessened challenges with description or testing quality (\textbf{C7}, \textbf{C3}). Interviewees (e.g., P18) suggested AI may have increased the number of PRs that have these artifacts at all.
% , e.g., \qualquote{Before AI\ldots 
% people hardly wrote anything.
% But now people write a lot in descriptions. Sometimes not very useful stuff, but
% at least they do write stuff} (P18).
% Future work could examine how accumulated AI-generated artifacts affect codebase debt over time, compared to their absence.
% Interviewees also
% noted that AI-generated tests can overfit or be vacuous or redundant
% (commonly-surfaced issues~\cite{}), e.g., \qualquote{AI is programmed to write tests that
% pass, right? And so it can be misleading and deceptive} (P16).

\section{Discussion and Conclusion}
\label{discussion}
We (1) situate review challenges, strategies, and outcomes with respect to prior work and (2) provide concrete recommendations for relevant stakeholders based on our results and concluding remarks.
% Developers found strategies which push review work upstream helpful to reduce asymmetric review burden and free up reviewers up for higher-level review. 
% As authors review more AI-generated code themselves and reviewers review for high-level concerns like design and intent, the roles of authors, reviewers, and AI review tooling may blur, suggesting that terms like ``author,'' ``reviewer,'' and even ``code review'' may no longer describe the work involved. 
% the above is like, really cool, but i think i'll probably deprioritize naming things as a contribution

% Below, we supply synthesized recommendations for  stakeholders based on our results.
\subsection{Comparison to Pre-AI Work on Code Review}
\label{comparisontopreai}
% Some challenges reported by our participants have been reported in work prior to AI adoption in software engineering. In a 2016 survey at Microsoft about salient code review challenges, \textit{review latency}, \textit{changeset size}, \textit{understanding the code's purpose}, \textit{understanding the motivations for a change}, and \textit{understanding how the change was implemented} were among the most frequent~\cite{}. Our survey demonstrates that these challenges have increased in frequency (see Section~\ref{}) and our study surfaces new AI-driven review challenges, overreliance on AI reviewing tools, shallow review, new PR quality issues, and productivity pressures. We also identify an erosion of nontechnical review benefits (see Section~\ref{}) and reviewer fatigue and lessening patience or focus induced by PR volume %hai

Our results demonstrate that many code review challenges identified in prior work before
AI adoption, including understanding a change and its motivation, review latency, review size, understanding change provenance, and communication
overhead~\cite{bacchelli2013expectations, macleod2017code}, all persist, but AI-assisted
authoring has made some more severe (e.g., PR volume and size) and others less so (e.g., untested changes). We also identified challenges without pre-AI
analogues, including a lack of author stewardship, uncertain intent, subtly incorrect code, missed reuse opportunities, and productivity
pressures. Similarly, our results demonstrate that established strategies (e.g., small changes,
clear descriptions, pre-review checks, and decision
traceability~\cite{macleod2017code}) have become more salient as ways to address
AI-specific challenges. Respondents also proposed new ones, such as documenting
implicit knowledge for AI and spending more resources on code review.
Review goals and outcomes are also changing. Knowledge transfer is a central outcome of review~\cite{bacchelli2013expectations}, but respondents suggested learning programming techniques through review matters less, while maintaining a mental model of the codebase through review still matters, though AI-specific review challenges make it harder.

% \textit{Challenges:} 
% (B\&B): understanding a change and its motivation and tools don't provide correct info (+1 from macleod)
% (macleod): receiving timely feedback + time constraints/latency, review size, understanding how a change was made i.e. provenance, managing multiple communication channels + identifying whom to talk to
% me: all of the above are shared, but we demonstrate AI adoption for authoring has made them more or less severe. we also identify new challenges w.r.t. before ai: lack of author stewardship and uncertain intent/unnecessary changes,  subtly incorrect code, lack of code reuse opportunities, productivity pressures 
% --> subpoint, about shift in outcomes -- in regard to knowledge transfer, learning programming techniques no longer important, but monitoring the codebase still is 
% --> subpoint, some challenges are not worsened now, e.g., bikeshedding (in fact quite the opposite)

% \textit{Strategies:} 
% (macleod): small/incremental changes, describe change motivations and what they do, test or analyze before review/check changes before review, set aside dedicated/bounded time, use tools that provide decision traceability

% me: all of the above are shared, but we demonstrate AI adoption for authoring has made them more salient and how they can target challenges from ai-assisted authoring. additionally, new strategies e.g. improve documentation of implicit knowledge and reports that organizations promoting and devoting more resources to CR would be helpful

\subsection{Recommendations for Stakeholders and Conclusion}
% Below, we supply synthesized recommendations for  stakeholders based on our results. \fixme{maybe include finding IDs by italicized recommendations so i don't have to include a sentence saying ``and here's our evidence''?}

% Interviewees (e.g., \fixme{pid}) discussed a \textit{proof of presence} feature that ensures human involvement in review sign-offs, but others indicated this feature might not go far enough in terms of getting human eyes on the PRs: \qualquote{}

% P24: \qualquote{I think that we should add some. some sort of gazing on a PR that ensures that a human has looked at it. So they do have a proof of presence, right, on the Prs. But the author just does that, right? And like, you know, and then that's done. And if nobody else commits, that's just going to ship, right? I think we need to kind of enforce more human. I'm not sure of the how yet, because I haven't had enough time to think about it. But like we need to try and enforce some sort of gating that ensures that Prs have to be looked at by a human in so much way, right? So they can't just be reviewed by AI and approved by AI, right? And that you need to have, like even if it's like, do you know when you say like, for example, when you look at it like a terms and conditions and you have to scroll to the bottom before you can agree.
% Right, like that, that I know that's just a very simplistic thing, but... Like that, you know, like, like, you know, even anything that shows that you that that you've actually looked through something from top to bottom}
\textit{\textbf{Organizations}: resource and incentivize code review.} Reflecting a broader shift in developer effort toward validation of AI-generated outputs~\cite{afroz2025fast}, participants reported spending more time on code review (\textbf{C9}, \textbf{AS2}) and wanted organizations to incentivize and allocate more resources to review (\textbf{S4}, \textbf{S8}, \textbf{S9}, \textbf{S12}). Organizations could make review contributions visible to teams and cite them in performance evaluations.
Interviewees (e.g., P14) lacked time to maintain AI review skills themselves and reported waiting for off-the-shelf tooling, suggesting organizations could fund dedicated ``context management’’ teams to maintain review skills tailored to code areas (\textbf{S3}).

\textit{\textbf{Organizations}: focus on multidimensional measures of code and review quality.} Participants indicated that productivity pressures could exacerbate shallow review and debt (see Section~\ref{r4aichal:productivitypressures}) and thought quality metrics were helpful (\textbf{S11}). Targeting metrics such as lines of code, review latency, and PR count could create perverse incentives~\cite{mattson2021measure, ko2019we}, and AI makes these metrics trivial to game. Measurements of developer productivity may be more effective if they are multidimensional, transparent, and target value beyond individual developer activity~\cite{umarji2009measuring, forsgren2021space}. 
% Although nontechnical review benefits are difficult to measure~\cite{caulo2020knowledge}, 
Measurements can also gauge the distribution of familiarity for regions of a codebase (i.e., the ``bus factor'')~\cite{thongtanunam2016revisiting, avelino2016novel} to determine which are particularly under threat of cognitive debt.

\textit{\textbf{Review workflows}: require authors to review PRs with AI
before human review.} The \textbf{highest-ranked} strategy in our study
was a pre-review step (\textbf{AIS1}, \textbf{S1}) that may clear up quality issues and
help authors understand and steward their PRs. Verifying
AI review findings took effort, adding to reviewer burden (\textbf{AC11}, \textbf{AC12}).
Requiring authors to do rounds of review with AI could reduce this effort for reviewers, improve author understanding, reduce author misevaluation
of AI comments (\textbf{AC4}, \textbf{AC13}), create documented feedback to improve AI
tooling (\textbf{AIS12}), and clear up low-level issues before human review. However, authors and reviewers have different perspectives, so author-run AI review may not
by itself free reviewers up to look for higher-level concerns.
Since developers saw duplicate or previously addressed AI comments
(\textbf{AC10}, \textbf{AC15}), tools could be iteration-aware (e.g., via
fingerprinting~\cite{spacco2006tracking} or semantic similarity
detection~\cite{ammu2025amalgamation}). Assigning reviewers only after PRs are ready to merge and AI comments are addressed could also ease review routing and notification fatigue (\textbf{S5}, \textbf{S7}).

\textit{\textbf{Tools}: prioritize reviewer attention between and
within PRs.} Developers prioritized their attention between and within PRs (\textbf{AS1--8}) by importance or ease of
understanding. Current review tools were not built for today's PR volume,
scope, and size, leading some interviewees (e.g., P1) to build their own.
Tools could direct attention to risky or high-priority PRs and issues (e.g.,
RADAR~\cite{adams2026automating}), and route reviews by author request, by the
concerns humans are best positioned to address, or for developer awareness.
However, directing attention may bias reviewers and reduce ability to independently identify defects~\cite{fregnan2022first, tufano2025deep}, so tools could mitigate this
(e.g., by signaling uncertainty~\cite{vasconcelos2025generation}).

\textit{\textbf{Tools}: surface author intent, author understanding, and PR
provenance.} Developers now review for high-level concerns (\textbf{RC1}),
but review tools center the diff as the main artifact, a mismatch between the \textit{unit of
analysis} tools surface and the \textit{unit of attention} review
demands~\cite{soderberg2022understanding}. Developers found strategies
surfacing intent, understanding, and provenance helpful (\textbf{S2}, \textbf{S10}).
Tools could show ``heatmaps'' of AI and human authoring activity (e.g., via
fine-grained interaction data~\cite{ziegler2022productivity} and its
visualization~\cite{hill1992edit, eick1992seesoft}), addressing both the
desire for provenance information and the difficulty of labeling PRs as AI- or
human-authored. Tools could also make the author's intent or
design a primary review artifact (as practiced by P22 and
in~\cite{maddila2026code}), while keeping code accessible, since
interviewees still inspected it (e.g., to verify AI comments). 

While our study surfaces challenges from AI-assisted authoring, the same advances that are disrupting code review now may ultimately help resolve them in the future. We believe our results demonstrate that AI could strengthen value from human review, rather than undermining it.

\section*{Data Availability}
Because this study was conducted within a proprietary organization, data cannot be publicly released. In particular, we cannot share individual survey responses or interview transcripts due to confidentiality and privacy constraints. To support reproducibility, we provide the survey instrument, interview protocol, and aggregated survey results as supplemental material. This approach is consistent with prior software engineering research conducted in industrial settings.

%%
%% The next two lines define the bibliography style to be used, and % the
%bibliography file.
\bibliographystyle{ACM-Reference-Format}
\bibliography{code_review}


\section{Tables}

\begin{table}[h]
\centering
\caption{Difficulty keeping up with code review volume and pace over the past 6 weeks.}
\label{tab:volume-difficulty-summary}
\begingroup
\fontsize{7.2}{8.5}\selectfont
\setlength{\tabcolsep}{0pt}
\setlength{\aboverulesep}{0.2ex}
\setlength{\belowrulesep}{0.2ex}
\renewcommand{\arraystretch}{1.12}
\sbox{\KaiaTableBox}{%
\begin{tabular}{@{}>{\raggedright\arraybackslash}p{0.06\columnwidth}@{\hspace{6pt}}>{\raggedright\arraybackslash}p{0.38\columnwidth}@{\hspace{6pt}}>{\raggedright\arraybackslash}p{0.04\columnwidth}@{\hspace{6pt}}>{\raggedleft\arraybackslash}p{0.08\columnwidth}@{\hspace{6pt}}>{\raggedright\arraybackslash}p{\dimexpr0.38\columnwidth-6pt\relax}@{}}
\toprule
\textbf{ID} & \textbf{Challenge} & \textbf{Cat} & \textbf{Count} & \multicolumn{1}{@{\hspace{6pt}}p{\dimexpr0.38\columnwidth-6pt\relax}@{}}{\KaiaDistributionHeader{Distribution}} \\\\
\midrule
V1 & To what extent do you agree or disagree with the following statement: Over the past 6 weeks, keeping up with the volume and pace of code I was expected or wanted to review was difficult. & X & 911 & \KaiaFrequencyBar{0.0450054885}{0.1394072448}{0.1931942920}{0.3929747530}{0.2294182217}{18}{62} \\\\
\bottomrule
\end{tabular}%
}
\usebox{\KaiaTableBox}
\par\vspace{0pt}
\vspace{0.2ex}
\begin{minipage}{\wd\KaiaTableBox}
\fontsize{6.5}{7.5}\selectfont
\KaiaLegend{KaiaFreqOne}{Strongly disagree}\quad\KaiaLegend{KaiaFreqTwo}{Somewhat disagree}\quad\KaiaLegend{KaiaFreqThree}{Neither agree nor disagree}\quad\KaiaLegend{KaiaFreqFour}{Somewhat agree}\quad\KaiaLegend{KaiaFreqFive}{Strongly agree}
\par\vspace{0.2ex}
\noindent\rule{\linewidth}{0.3pt}
\end{minipage}
\endgroup
\end{table}

\begin{table}[t]
\centering
\caption{Attention challenge comparisons before vs after AI involvement, ordered by higher 'more often' responses. Respondents per item ranged from n=315 to n=317.}
\label{tab:attention-comparison-summary}
\begingroup
\fontsize{6.6}{7.6}\selectfont
\setlength{\tabcolsep}{0pt}
\setlength{\aboverulesep}{0.2ex}
\setlength{\belowrulesep}{0.2ex}
\renewcommand{\arraystretch}{1.10}
\newcommand{\KaiaBipolarBar}[7]{%
  \begingroup%
  \def\KaiaColorOne{KaiaCompFive}%
  \def\KaiaColorTwo{KaiaCompFour}%
  \def\KaiaColorThree{KaiaCompThree}%
  \def\KaiaColorFour{KaiaCompTwo}%
  \def\KaiaColorFive{KaiaCompOne}%
  \KaiaDistributionBar{#1}{#2}{#3}{#4}{#5}{#6}{#7}%
  \endgroup%
}
\sbox{\KaiaTableBox}{%
\begin{tabular}{@{}>{\raggedright\arraybackslash}p{0.05\columnwidth}@{\hspace{4pt}}>{\raggedright\arraybackslash}p{0.25\columnwidth}@{\hspace{4pt}}>{\raggedright\arraybackslash}p{0.03\columnwidth}@{\hspace{4pt}}>{\raggedleft\arraybackslash}p{0.06\columnwidth}@{\hspace{4pt}}>{\raggedright\arraybackslash}p{0.27\columnwidth}@{\hspace{4pt}}>{\raggedleft\arraybackslash}p{0.09\columnwidth}@{\hspace{4pt}}>{\raggedleft\arraybackslash}p{0.08\columnwidth}@{\hspace{4pt}}>{\raggedleft\arraybackslash}p{0.09\columnwidth}@{}}
\toprule
\textbf{ID} & \textbf{Challenge} & \textbf{Cat} & \textbf{Count} & \multicolumn{1}{@{\hspace{4pt}}p{0.27\columnwidth}@{}}{\KaiaDistributionHeader{Distribution}} & \textbf{More vs Less} & \textbf{p} & \textbf{p adj.} \\\\
\midrule
C1 & A PR I reviewed had a changeset that was too large or touched too many files & X & 316 & \KaiaBipolarBar{0.0348101266}{0.0949367089}{0.2658227848}{0.3797468354}{0.2246835443}{13}{60} & 60\% vs 13\% & $p < .001$ & $p < .001$ \\
C4 & I didn't do as thorough of a review as I would have liked to & X & 315 & \KaiaBipolarBar{0.0761904762}{0.1174603175}{0.3333333333}{0.3619047619}{0.1111111111}{19}{47} & 47\% vs 19\% & $p < .001$ & $p < .001$ \\
C2 & I felt pressure to review faster & X & 317 & \KaiaBipolarBar{0.0473186120}{0.0851735016}{0.4447949527}{0.3154574132}{0.1072555205}{13}{42} & 42\% vs 13\% & $p < .001$ & $p < .001$ \\
C9 & My reviewing responsibilities affected my ability to complete other job responsibilities & X & 317 & \KaiaBipolarBar{0.0662460568}{0.1482649842}{0.4668769716}{0.2429022082}{0.0757097792}{21}{32} & 32\% vs 21\% & $p = .06$ & $p = .10$ \\
C18 & I needed to repeat feedback I'd already given on other PRs & X & 316 & \KaiaBipolarBar{0.0886075949}{0.1234177215}{0.5537974684}{0.1708860759}{0.0632911392}{21}{23} & 23\% vs 21\% & $p = .80$ & $p = .80$ \\
C19 & A PR I wanted to review was merged without my review & X & 315 & \KaiaBipolarBar{0.0507936508}{0.0920634921}{0.6666666667}{0.1587301587}{0.0317460317}{14}{19} & 19\% vs 14\% & $p = .63$ & $p = .73$ \\
C17 & A PR I wanted to review did not need my review when I looked at it (e.g., it was already approved, wasn't ready, did not fit my expertise) & X & 316 & \KaiaBipolarBar{0.0506329114}{0.0917721519}{0.7278481013}{0.1107594937}{0.0189873418}{14}{13} & 13\% vs 14\% & $p = .20$ & $p = .28$ \\
\bottomrule
\end{tabular}%
}
\KaiaCheckTableWidth
\usebox{\KaiaTableBox}
\par\vspace{0pt}
\vspace{0.2ex}
\begin{minipage}{\wd\KaiaTableBox}
\fontsize{6.2}{7.2}\selectfont
\KaiaLegend{KaiaCompFive}{Much less often}\quad\KaiaLegend{KaiaCompFour}{Less often}\quad\KaiaLegend{KaiaCompThree}{About the same}\quad\KaiaLegend{KaiaCompTwo}{More often}\quad\KaiaLegend{KaiaCompOne}{Much more often}
\par\vspace{0.2ex}
\noindent\rule{\linewidth}{0.3pt}
\end{minipage}
\endgroup
\end{table}

\begin{table}[t]
\centering
\caption{Volume and pace comparison before vs after AI involvement ($n=825$). Respondents were asked: \textit{``Compared to before PR authors used AI for code generation, keeping up with the volume and pace of code I'm expected or want to review is \_\_\_\_.''}}
\label{tab:volume-comparison-summary}
\begingroup
\fontsize{7.5}{8.8}\selectfont
\setlength{\tabcolsep}{4pt}
\setlength{\aboverulesep}{0.2ex}
\setlength{\belowrulesep}{0.2ex}
\renewcommand{\arraystretch}{1.05}
\newcommand{\KaiaBipolarBar}[7]{%
  \begingroup%
  \def\KaiaColorOne{KaiaCompFive}%
  \def\KaiaColorTwo{KaiaCompFour}%
  \def\KaiaColorThree{KaiaCompThree}%
  \def\KaiaColorFour{KaiaCompTwo}%
  \def\KaiaColorFive{KaiaCompOne}%
  \KaiaDistributionBar{#1}{#2}{#3}{#4}{#5}{#6}{#7}%
  \endgroup%
}
\sbox{\KaiaTableBox}{%
\begin{tabular}{@{}>{\raggedleft\arraybackslash}p{0.09\columnwidth}@{\hspace{6pt}}>{\raggedright\arraybackslash}p{0.42\columnwidth}@{\hspace{6pt}}>{\raggedright\arraybackslash}p{0.22\columnwidth}@{\hspace{6pt}}>{\raggedleft\arraybackslash}p{0.08\columnwidth}@{\hspace{6pt}}>{\raggedleft\arraybackslash}p{0.08\columnwidth}@{}}
\toprule
\textbf{Count} & \multicolumn{1}{@{}p{0.42\columnwidth}@{}}{\KaiaDistributionHeader{Distribution}} & \textbf{More difficult vs Less} & \textbf{p} & \textbf{p adj.} \\
\midrule
825 & \KaiaBipolarBar{0.0472727273}{0.1745454545}{0.2206060606}{0.3927272727}{0.1648484848}{22}{56} & 56\% vs 22\% & $p<.001$ & $p<.001$ \\
\bottomrule
\end{tabular}%
}
\KaiaCheckTableWidth
\usebox{\KaiaTableBox}
\vspace{0.2ex}
\begin{minipage}{\wd\KaiaTableBox}
\fontsize{6.5}{7.5}\selectfont
\KaiaLegend{KaiaCompFive}{Much easier}\quad\KaiaLegend{KaiaCompFour}{Easier}\quad\KaiaLegend{KaiaCompThree}{About the same}\quad\KaiaLegend{KaiaCompTwo}{More difficult}\quad\KaiaLegend{KaiaCompOne}{Much more difficult}
\par\vspace{0.2ex}
\noindent\rule{\linewidth}{0.3pt}
\end{minipage}
\endgroup
\end{table}

\begin{table}[t]
\centering
\caption{PR quality challenge comparisons before vs after AI involvement, ordered by higher 'more often' responses. Respondents per item ranged from n=308 to n=308.}
\label{tab:pr-quality-comparison-summary}
\begingroup
\fontsize{6.6}{7.6}\selectfont
\setlength{\tabcolsep}{0pt}
\setlength{\aboverulesep}{0.2ex}
\setlength{\belowrulesep}{0.2ex}
\renewcommand{\arraystretch}{1.10}
\newcommand{\KaiaBipolarBar}[7]{%
  \begingroup%
  \def\KaiaColorOne{KaiaCompFive}%
  \def\KaiaColorTwo{KaiaCompFour}%
  \def\KaiaColorThree{KaiaCompThree}%
  \def\KaiaColorFour{KaiaCompTwo}%
  \def\KaiaColorFive{KaiaCompOne}%
  \KaiaDistributionBar{#1}{#2}{#3}{#4}{#5}{#6}{#7}%
  \endgroup%
}
\sbox{\KaiaTableBox}{%
\begin{tabular}{@{}>{\raggedright\arraybackslash}p{0.05\columnwidth}@{\hspace{4pt}}>{\raggedright\arraybackslash}p{0.25\columnwidth}@{\hspace{4pt}}>{\raggedright\arraybackslash}p{0.03\columnwidth}@{\hspace{4pt}}>{\raggedleft\arraybackslash}p{0.06\columnwidth}@{\hspace{4pt}}>{\raggedright\arraybackslash}p{0.27\columnwidth}@{\hspace{4pt}}>{\raggedleft\arraybackslash}p{0.09\columnwidth}@{\hspace{4pt}}>{\raggedleft\arraybackslash}p{0.08\columnwidth}@{\hspace{4pt}}>{\raggedleft\arraybackslash}p{0.09\columnwidth}@{}}
\toprule
\textbf{ID} & \textbf{Challenge} & \textbf{Cat} & \textbf{Count} & \multicolumn{1}{@{\hspace{4pt}}p{0.27\columnwidth}@{}}{\KaiaDistributionHeader{Distribution}} & \textbf{More vs Less} & \textbf{p} & \textbf{p adj.} \\\\
\midrule
C10 & Code that appeared correct on the surface, but was wrong in a way that was difficult to notice & X & 308 & \KaiaBipolarBar{0.0811688312}{0.1753246753}{0.3409090909}{0.2857142857}{0.1168831169}{26}{40} & 40\% vs 26\% & $p = .007$ & $p = .02$ \\
C11 & Code that reimplemented something that exists somewhere else & X & 308 & \KaiaBipolarBar{0.0941558442}{0.1688311688}{0.3506493506}{0.2792207792}{0.1071428571}{26}{39} & 39\% vs 26\% & $p = .05$ & $p = .09$ \\
C12 & A PR addressed something that did not seem important or necessary & X & 308 & \KaiaBipolarBar{0.0941558442}{0.1525974026}{0.3733766234}{0.2889610390}{0.0909090909}{25}{38} & 38\% vs 25\% & $p = .07$ & $p = .09$ \\
C3 & PR lacked evidence that it was adequately tested or verified & X & 308 & \KaiaBipolarBar{0.1006493506}{0.2435064935}{0.3116883117}{0.2337662338}{0.1103896104}{34}{34} & 34\% vs 34\% & $p = .85$ & $p = .85$ \\
C6 & Code that didn't follow architectural or stylistic conventions in the codebase & X & 308 & \KaiaBipolarBar{0.1331168831}{0.2110389610}{0.3279220779}{0.2435064935}{0.0844155844}{34}{33} & 33\% vs 34\% & $p = .25$ & $p = .29$ \\
C15 & The same kinds of mistakes appeared across many different PRs & X & 308 & \KaiaBipolarBar{0.1103896104}{0.2142857143}{0.4350649351}{0.1850649351}{0.0551948052}{32}{24} & 24\% vs 32\% & $p = .01$ & $p = .03$ \\
C7 & PR description or title did not give me context I needed to review effectively & X & 308 & \KaiaBipolarBar{0.1785714286}{0.2954545455}{0.3279220779}{0.1103896104}{0.0876623377}{47}{20} & 20\% vs 47\% & $p < .001$ & $p < .001$ \\
\bottomrule
\end{tabular}%
}
\KaiaCheckTableWidth
\usebox{\KaiaTableBox}
\par\vspace{0pt}
\vspace{0.2ex}
\begin{minipage}{\wd\KaiaTableBox}
\fontsize{6.2}{7.2}\selectfont
\KaiaLegend{KaiaCompFive}{Much less often}\quad\KaiaLegend{KaiaCompFour}{Less often}\quad\KaiaLegend{KaiaCompThree}{About the same}\quad\KaiaLegend{KaiaCompTwo}{More often}\quad\KaiaLegend{KaiaCompOne}{Much more often}
\par\vspace{0.2ex}
\noindent\rule{\linewidth}{0.3pt}
\end{minipage}
\endgroup
\end{table}

\begin{table}[t]
\centering
\caption{Social interaction challenge comparisons before vs after AI involvement, ordered by higher 'more often' responses. Respondents per item ranged from n=242 to n=292.}
\label{tab:social-interactions-comparison-summary}
\begingroup
\fontsize{6.6}{7.6}\selectfont
\setlength{\tabcolsep}{0pt}
\setlength{\aboverulesep}{0.2ex}
\setlength{\belowrulesep}{0.2ex}
\renewcommand{\arraystretch}{1.10}
\newcommand{\KaiaBipolarBar}[7]{%
  \begingroup%
  \def\KaiaColorOne{KaiaCompFive}%
  \def\KaiaColorTwo{KaiaCompFour}%
  \def\KaiaColorThree{KaiaCompThree}%
  \def\KaiaColorFour{KaiaCompTwo}%
  \def\KaiaColorFive{KaiaCompOne}%
  \KaiaDistributionBar{#1}{#2}{#3}{#4}{#5}{#6}{#7}%
  \endgroup%
}
\sbox{\KaiaTableBox}{%
\begin{tabular}{@{}>{\raggedright\arraybackslash}p{0.05\columnwidth}@{\hspace{4pt}}>{\raggedright\arraybackslash}p{0.25\columnwidth}@{\hspace{4pt}}>{\raggedright\arraybackslash}p{0.03\columnwidth}@{\hspace{4pt}}>{\raggedleft\arraybackslash}p{0.06\columnwidth}@{\hspace{4pt}}>{\raggedright\arraybackslash}p{0.27\columnwidth}@{\hspace{4pt}}>{\raggedleft\arraybackslash}p{0.09\columnwidth}@{\hspace{4pt}}>{\raggedleft\arraybackslash}p{0.08\columnwidth}@{\hspace{4pt}}>{\raggedleft\arraybackslash}p{0.09\columnwidth}@{}}
\toprule
\textbf{ID} & \textbf{Challenge} & \textbf{Cat} & \textbf{Count} & \multicolumn{1}{@{\hspace{4pt}}p{0.27\columnwidth}@{}}{\KaiaDistributionHeader{Distribution}} & \textbf{More vs Less} & \textbf{p} & \textbf{p adj.} \\\\
\midrule
C14 & When I asked about a change, the author could not explain or justify it & X & 282 & \KaiaBipolarBar{0.1560283688}{0.1028368794}{0.2553191489}{0.3085106383}{0.1773049645}{26}{49} & 49\% vs 26\% & $p = .01$ & $p = .03$ \\
C5 & I had to spend additional time determining how or why changes worked because the author had not adequately explained them & X & 292 & \KaiaBipolarBar{0.0924657534}{0.1609589041}{0.3287671233}{0.3047945205}{0.1130136986}{25}{42} & 42\% vs 25\% & $p = .01$ & $p = .03$ \\
C8 & I found problems that the author should reasonably have caught before review & X & 291 & \KaiaBipolarBar{0.0893470790}{0.1890034364}{0.3745704467}{0.2749140893}{0.0721649485}{28}{35} & 35\% vs 28\% & $p = .50$ & $p = .50$ \\
C13 & I lacked information about the author’s intent or reasoning that I needed to review effectively & X & 285 & \KaiaBipolarBar{0.1157894737}{0.2175438596}{0.3543859649}{0.2526315789}{0.0596491228}{33}{31} & 31\% vs 33\% & $p = .17$ & $p = .24$ \\
C16 & I experienced friction with authors or other developers because of differing expectations about code review (e.g., its thoroughness, scope, or speed) & X & 262 & \KaiaBipolarBar{0.1106870229}{0.1221374046}{0.5190839695}{0.1870229008}{0.0610687023}{23}{25} & 25\% vs 23\% & $p = .41$ & $p = .47$ \\
C20 & My comments were met with responses or changes that did not address my concerns & X & 275 & \KaiaBipolarBar{0.1272727273}{0.1781818182}{0.4545454545}{0.1890909091}{0.0509090909}{31}{24} & 24\% vs 31\% & $p = .01$ & $p = .03$ \\
C21 & I felt that the author was not being respectful of my time & X & 242 & \KaiaBipolarBar{0.1694214876}{0.1446280992}{0.4669421488}{0.1198347107}{0.0991735537}{31}{22} & 22\% vs 31\% & $p = .03$ & $p = .04$ \\
\bottomrule
\end{tabular}%
}
\KaiaCheckTableWidth
\usebox{\KaiaTableBox}
\par\vspace{0pt}
\vspace{0.2ex}
\begin{minipage}{\wd\KaiaTableBox}
\fontsize{6.2}{7.2}\selectfont
\KaiaLegend{KaiaCompFive}{Much less often}\quad\KaiaLegend{KaiaCompFour}{Less often}\quad\KaiaLegend{KaiaCompThree}{About the same}\quad\KaiaLegend{KaiaCompTwo}{More often}\quad\KaiaLegend{KaiaCompOne}{Much more often}
\par\vspace{0.2ex}
\noindent\rule{\linewidth}{0.3pt}
\end{minipage}
\endgroup
\end{table}

\begin{table}[t]
\centering
\caption{\textbf{AI review challenges}, ordered by the percentage answering about half the time or more. Respondents per item ranged from $n=963-1214$. Categories: ST = social tension, UE = usage effort, OR = overreliance, PM = perception management, PR = precision, PS = presentation.}
\label{tab:ai-review-challenge-summary}
\begingroup
\fontsize{7.2}{8.0}\selectfont
\setlength{\tabcolsep}{0pt}
\setlength{\aboverulesep}{0.1ex}
\setlength{\belowrulesep}{0.1ex}
\renewcommand{\arraystretch}{0.92}
\sbox{\KaiaTableBox}{%
\begin{tabular}{@{}>{\raggedright\arraybackslash}p{0.06\columnwidth}@{\hspace{6pt}}>{\raggedright\arraybackslash}p{0.45\columnwidth}@{\hspace{6pt}}>{\raggedright\arraybackslash}p{0.02\columnwidth}@{\hspace{6pt}}>{\raggedright\arraybackslash}p{\dimexpr0.39\columnwidth-6pt\relax}@{}}
\toprule
\textbf{ID} & \textbf{Challenge} & \textbf{Cat.} & \multicolumn{1}{@{\hspace{6pt}}p{\dimexpr0.39\columnwidth-6pt\relax}@{}}{\KaiaDistributionHeader{Distribution}} \\\\
\midrule
AC1 & Not all important issues were caught by AI comments & PR & \KaiaFrequencyBar{0.0487603306}{0.4123966942}{0.2520661157}{0.2322314050}{0.0545454545}{46}{54} \\\\
AC2 & AI comments were overly verbose or complex & PS & \KaiaFrequencyBar{0.1267605634}{0.3405136703}{0.1748135874}{0.2336371168}{0.1242750621}{47}{53} \\\\
AC3 & After relying on AI review, I worried that I had not personally reviewed the PR closely enough & OR & \KaiaFrequencyBar{0.2242798354}{0.3127572016}{0.1759259259}{0.1820987654}{0.1049382716}{54}{46} \\\\
AC4 & AI-generated comments were not engaged with by the author & PM & \KaiaFrequencyBar{0.1489184692}{0.3968386023}{0.2212978369}{0.2121464226}{0.0207986689}{55}{45} \\\\
AC5 & An AI-generated comment was not worth leaving (e.g., was a nitpick) & PR & \KaiaFrequencyBar{0.0620860927}{0.5198675497}{0.2657284768}{0.1365894040}{0.0157284768}{58}{42} \\\\
AC6 & AI comments were too numerous & PS & \KaiaFrequencyBar{0.2230639731}{0.3872053872}{0.1784511785}{0.1473063973}{0.0639730640}{61}{39} \\\\
AC7 & An AI-generated comment was inaccurate or uncomplete & PR & \KaiaFrequencyBar{0.0477759473}{0.6037891269}{0.2413509061}{0.0963756178}{0.0107084020}{65}{35} \\\\
AC8 & AI review gave incorrect feedback because it lacked unwritten project knowledge & PR & \KaiaFrequencyBar{0.1406570842}{0.5410677618}{0.1755646817}{0.1232032854}{0.0195071869}{68}{32} \\\\
AC9 & AI review gave incorrect feedback because it did not account for relevant code or documentation beyond the PR & PR & \KaiaFrequencyBar{0.1204081633}{0.5948979592}{0.1704081633}{0.0989795918}{0.0153061224}{72}{28} \\\\
AC10 & An AI-generated comment made the same kind of mistake I saw AI make in the past & PR & \KaiaFrequencyBar{0.2162393162}{0.5025641026}{0.1487179487}{0.1153846154}{0.0170940171}{72}{28} \\\\
AC11 & AI review feedback was difficult or time-consuming to verify & UE & \KaiaFrequencyBar{0.2654320988}{0.4578189300}{0.1676954733}{0.0895061728}{0.0195473251}{72}{28} \\\\
AC12 & Providing necessary context or configuration for AI review took too much effort & UE & \KaiaFrequencyBar{0.3312629400}{0.4089026915}{0.1501035197}{0.0910973085}{0.0186335404}{74}{26} \\\\
AC13 & Some AI-generated comments that I thought should be responded to did not have author engagement & PM & \KaiaFrequencyBar{0.3053691275}{0.4681208054}{0.1367449664}{0.0780201342}{0.0117449664}{77}{23} \\\\
AC14 & PR authors expressed frustration with AI comments & ST & \KaiaFrequencyBar{0.4104289319}{0.4003364172}{0.1135407906}{0.0630782170}{0.0126156434}{81}{19} \\\\
AC15 & AI review repeated a mistake despite receiving prior instructions not to & PR & \KaiaFrequencyBar{0.3966770509}{0.4361370717}{0.1079958463}{0.0508826584}{0.0083073728}{83}{17} \\\\
AC16 & AI comments repeated points from existing comments & PS & \KaiaFrequencyBar{0.3921074727}{0.4500419815}{0.0915197313}{0.0528967254}{0.0134340890}{84}{16} \\\\
AC17 & AI comments were left in a broken or hard-to-read format & PS & \KaiaFrequencyBar{0.6141666667}{0.2691666667}{0.0500000000}{0.0491666667}{0.0175000000}{88}{12} \\\\
\bottomrule
\end{tabular}%
}
\KaiaCheckTableWidth
\usebox{\KaiaTableBox}
\par\vspace{0pt}
\begin{minipage}{\wd\KaiaTableBox}
\fontsize{6.2}{7.0}\selectfont
\KaiaLegend{KaiaFreqOne}{Never}\quad\KaiaLegend{KaiaFreqTwo}{Sometimes}\quad\KaiaLegend{KaiaFreqThree}{About half the time}\quad\KaiaLegend{KaiaFreqFour}{Most of the time}\quad\KaiaLegend{KaiaFreqFive}{Always}\\[-4pt]
\noindent\rule{\linewidth}{\heavyrulewidth}
\end{minipage}
\endgroup
\end{table}

\begin{table}[t]
\captionsetup{font=scriptsize, labelfont=scriptsize}
\caption{\textbf{AI review tool strategies, ranked by perceived helpfulness.} $n=1203$ respondents indicated whether they used each strategy; 3\% selected Other, 3\% selected ``None, despite challenges,'' and 2\% selected ``None, works off-the-shelf.'' $n=942$ respondents (those whose top favorites were not Other) were used to fit a Plackett-Luce model to rank strategies, which estimates each strategy's relative worth from respondents' top-choice preferences among used strategies. Categories: ST = social tension, UE = usage effort, OR = overreliance, PM = perception management, PR = precision, PS = presentation. Empty categories help with challenges from AI-assisted authoring.}
\label{tab:ai-strats}
\centering
\scriptsize
\setlength{\tabcolsep}{2pt}
\renewcommand{\arraystretch}{1.0}
\begin{tabularx}{\textwidth}{@{}l X l l l l@{}}
\toprule
ID & Option & Cat & Used \% & worth & CI \\
\midrule
\rowcolor{rowColor} AIS1 & Have authors use AI review on their PRs before submitting for human review & -- & 69\% & 0.29 & [0.22, 0.35] \\
AIS2 & Have AI review do multiple passes or put multiple agents on one PR & PR & 55\% & 0.20 & [0.14, 0.25] \\
\rowcolor{rowColor} AIS3 & Add repository- or team-specific review expectations to reusable instructions (e.g., a skill) & PR & 55\% & 0.11 & [0.08, 0.15] \\
AIS4 & Give AI review access to more relevant code or supporting information & PR & 46\% & 0.11 & [0.07, 0.15] \\
\rowcolor{rowColor} AIS5 & Improve documentation of the codebase or implicit knowledge to aid AI review & PR & 48\% & 0.08 & [0.05, 0.10] \\
AIS6 & Instruct AI review to address a few specific concerns rather than do a generic pass & PR & 47\% & 0.06 & [0.04, 0.08] \\
\rowcolor{rowColor} AIS7 & Automate AI review to run on a schedule or based on triggers (e.g., when a PR is submitted) & -- & 38\% & 0.03 & [0.02, 0.05] \\
AIS8 & Tune the language, tone, or emphasis of instructions (e.g., "do this at max effort," "go step by step") & PR & 34\% & 0.03 & [0.02, 0.05] \\
\rowcolor{rowColor} AIS9 & Customize AI review for different repos or areas (e.g., pointing it to different context, using a different model) & PR & 39\% & 0.03 & [0.02, 0.05] \\
AIS10 & Encourage developers to engage with AI review comments & PM & 48\% & 0.02 & [0.01, 0.02] \\
\rowcolor{rowColor} AIS11 & Scale back AI review based on negative developer feedback & ST & 11\% & 0.02 & [0.00, 0.03] \\
AIS12 & Refine AI review based on human feedback from review sessions & PR & 25\% & 0.01 & [0.01, 0.02] \\
\rowcolor{rowColor} AIS13 & Instruct AI review to adjust the length, tone, or format of comments & PS & 26\% & 0.01 & [0.00, 0.01] \\
AIS14 & Change how AI review findings are displayed to the user (e.g., in a dashboard or aggregated report, inline comments) & PS & 10\% & 0.00 & [0.00, 0.01] \\
\bottomrule
\end{tabularx}
\end{table}

% \usepackage{booktabs}
% \usepackage{graphicx}
% \usepackage[table]{xcolor}
\begin{table*}[]
\centering
\caption{\textbf{Interview participant table.} Rev. hrs/week = the number of hours doing code review per week, PRs/week = number of reviewed PRs per week, AI review use = how frequently they use AI for review. A ``monorepo'' is a large repository containing multiple codebases. Most data for P24 is missing from nonresponse.}
\resizebox{\textwidth}{!}{%
\begin{tabular}{@{}llllllllll@{}}
\toprule
ID  & Job Position        & YOE & Rev. hrs/week & PRs/week & AI review use & Country       & Age & Gender & Recruitment source \\ \midrule
\rowcolor{rowColor}
P1  & Engineering Manager & 15               & 6-10            & 40                & Always                  & United States & 36  & M      & Monorepo A         \\
P2  & Software Architect  & 16               & 20+             & --               & Rarely                  & --           & -- & M      & Monorepo A         \\
\rowcolor{rowColor}
P3  & Software Engineer   & 12               & 6-10            & 20-50             & Always                  & United States & 38  & W      & Monorepo B         \\
P4  & Software Engineer   & 16               & 6-10            & 100               & Never                   & United States & 38  & --    & Monorepo A         \\
\rowcolor{rowColor}
P5  & Software Engineer   & 14               & 1-2             & 30                & Never                   & United States & 36  & M      & Monorepo A         \\
P6  & Software Engineer   & 18               & 11-20           & 22                & Sometimes               & United States & 40  & M      & Monorepo A         \\
\rowcolor{rowColor}
P7  & Software Engineer   & 11               & 3-5             & 40                & Sometimes               & United States & 32  & W      & Monorepo B         \\
P8  & Software Engineer   & 5                & 6-10            & 15+               & Never                   & United States & 27  & W      & Monorepo B         \\
\rowcolor{rowColor}
P9  & Software Engineer   & 13               & 3-5             & 12                & Sometimes               & Norway        & 37  & M      & Monorepo B         \\
P10 & Software Engineer   & 15               & 6-10            & 10                & Sometimes               & United States & 36  & W      & Monorepo B         \\
\rowcolor{rowColor}
P11 & Software Engineer   & 8                & 11-20           & 10+               & Sometimes               & United States & 32  & M      & Monorepo A         \\
P12 & Software Engineer   & 7                & 3-5             & 10                & Never                   & United States & 26  & M      & Monorepo A         \\
\rowcolor{rowColor}
P13 & Software Architect  & 14               & 11-20           & 50                & Always                  & Croatia       & 37  & M      & Monorepo B         \\
P14 & Engineering Manager & 14               & 3-5             & 10                & Never                   & United States & 36  & M      & Monorepo A         \\
\rowcolor{rowColor}
P15 & Software Engineer   & 10               & 6-10            & 10                & Never                   & Brazil        & 30  & M      & Monorepo B         \\
P16 & Software Engineer   & 15+              & 11-20           & 15-30             & N/A                     & United States & 47  & M      & Monorepo A         \\
\rowcolor{rowColor}
P17 & Software Engineer   & 3                & 3-5             & 3                 & Always                  & United States & 27  & W      & Word of mouth      \\
P18 & Software Engineer   & 12               & 6-10            & 8                 & Always                  & India         & 35  & M      & Monorepo B         \\
\rowcolor{rowColor}
P19 & Software Engineer   & 26               & 6-10            & 10+               & Always                  & United States & 46  & M      & Pilot              \\
P20 & Engineering Manager & 26               & 3-5             & --               & Always                  & India         & 48  & M      & Repo Collection C  \\
\rowcolor{rowColor}
P21 & Software Architect  & 30               & 1-2             & 5                 & Always                  & United States & 57  & M      & Repo Collection C  \\
P22 & Engineering Manager & 30+              & 6-10            & 15+               & Always                  & United States & 54  & M      & Repo Collection C  \\
\rowcolor{rowColor}
P23 & Software Engineer   & 30               & 20+             & 50+               & Sometimes               & Australia     & 46  & M      & Repo Collection C  \\
P24 & Software Engineer   & --              & --             & --               & --                     & Ireland       & -- & --    & Repo Collection C  \\ \bottomrule
\end{tabular}%
}
\end{table*}

\end{document}